\documentclass[11pt,a4paper]{article}

\usepackage{amsmath,amssymb,tikz,hyperref,empheq}
\usepackage{graphicx}
 \usepackage{verbatim}
 \allowdisplaybreaks
\def\be{\begin{equation}}
\def\ee{\end{equation}}
\def\ba#1\ea{\begin{align}#1\end{align}}
\def\bg#1\eg{\begin{gather}#1\end{gather}}
\def\bm#1\em{\begin{multline}#1\end{multline}}
\def\bmd#1\emd{\begin{multlined}#1\end{multlined}}

\def\e{\epsilon}

\def\x{\xi}

\def\({\left(}
\def\){\right)}
\def\[{\left[}
\def\]{\right]}

\def \be {\begin{equation}}
\def \ee {\end{equation}}
\def \ba {\begin{array}}
\def \ea {\end{array}}
\def \bea{\begin{eqnarray}}
\def \eea{\end{eqnarray}}

\def \e {\epsilon}

\def\bea{\begin{eqnarray}}
\def\eea{\end{eqnarray}}

\newcommand{\bit}{\begin{itemize}}  \newcommand{\eit}{\end{itemize}}
\newcommand{\ben}{\begin{enumerate}}  \newcommand{\een}{\end{enumerate}}

\long\def\symbolfootnote[#1]#2{\begingroup%
\def\thefootnote{\fnsymbol{footnote}}\footnote[#1]{#2}\endgroup}

\newcommand{\sysu}{{\it School of Physics and Astronomy, Sun Yat-Sen University, 2 Daxue Road, Zhuhai 519082, China}}

\begin{document}
\thispagestyle{empty}
\begin{center}

~\vspace{20pt}

{\Large\bf Casimir Effect and Holographic Dual of Wedges}

\vspace{25pt}

Rong-Xin Miao ${}$\symbolfootnote[1]{Email:~\sf miaorx@mail.sysu.edu.cn}

\vspace{10pt}${}$\sysu

\vspace{2cm}

\begin{abstract}
This paper investigates the Casimir effect of a wedge and its holographic dual. We prove that the displacement operator universally determines the wedge Casimir effect in the smooth limit. Besides, we argue that the wedge Casimir energy increases with the opening angle and test it with several examples. Furthermore, we construct the holographic dual of wedges in AdS/BCFT in general dimensions. We verify that our proposal can produce the expected Casimir effect within smooth and singular limits. We observe that the Casimir energy density of a wedge increases with the brane tension. Next, we discuss the wedge contribution to holographic entanglement entropy and find it increases with the opening angle, similar to the wedge Casimir energy. Finally, we briefly discuss the holographic polygon in AdS$_3$/BCFT$_2$ and its generalization to higher dimensions. 
\end{abstract}

\end{center}

\newpage
\setcounter{footnote}{0}
\setcounter{page}{1}

\tableofcontents

\section{Introduction}

Casimir effect \cite{Casimir:1948dh} is a novel quantum effect that originates from the changes of zero-point energy due to the boundary or the background geometry. Casimir effect has been measured in experiments \cite{Mohideen:1998iz, Bressi:2002fr, Klimchitskaya:2009cw} and has wild applications in quantum field theory, condensed matter, and cosmology \cite{Plunien:1986ca, Bordag:2001qi, Milton:2004ya}. It is a candidate of dark energy \cite{Wang:2016och} and plays a vital role in worm-hole physics \cite{Morris:1988tu, Maldacena:2020sxe}.  

We are interested in the universal nature of the Casimir effect for boundary conformal field theories (BCFT) in flat wedge space
\begin{eqnarray}\label{metric} 
ds^2=dr^2+r^2 d\theta^2+\sum_{\hat{i},\hat{j}=1}^{d-2} \eta_{\hat{i}\hat{j}} dy^{\hat{i}} dy^{\hat{j}}, \ \ 0\le\theta\le \Omega,
\end{eqnarray}
where $ \eta_{\hat{i}\hat{j}} =\text{diag}(-1,1,...,1) $ and $\Omega$ is the opening angle of the wedge. See \cite{Antunes:2021qpy,Diatlyk:2024zkk, Bissi:2022bgu} for recent studies on the BCFT in a wedge. For simplicity, we impose the same conformal boundary conditions on the two boundaries $\theta=0$ and $\theta=\Omega$. As a result, there is only a geometrical defect at the corner of the wedge $r=0$. From the symmetries of the wedge and the vanishing of the trace and divergence of the renormalized stress tensor, we have generally \cite{Bissi:2022bgu, Deutsch:1978sc}
\begin{eqnarray}\label{Tij wedge0a} 
&&\langle T^{rr} \rangle=\frac{f(\theta,\Omega)-\frac{1}{(d-2)}\partial_{\theta}^2f(\theta,\Omega)}{r^d},\ \ \langle T^{\theta\theta} \rangle=\frac{-(d-1)f(\theta,\Omega)}{r^{d+2}},\\ \label{Tij wedge0b} 
&&\langle T^{r\theta} \rangle=\frac{-(d-1)\partial_{\theta}f(\theta,\Omega)}{(d-2)r^{d+1}}, \ \ \langle T^{\hat{i}\hat{j}} \rangle=\frac{f(\theta,\Omega)+\frac{1}{(d-2)^2}\partial_{\theta}^2f(\theta,\Omega)}{r^d} \eta^{\hat{i}\hat{j}}
\end{eqnarray}
where $f(\theta, \Omega)$ is a function vanishing when $\Omega=\pi$.  There are more symmetries for a ground state, where $\langle T^{r\theta} \rangle$ is expected to vanish, and $f(\theta, \Omega)$ becomes $\theta$-independent. Thus, it is expected for a ground state \cite{Deutsch:1978sc}
\begin{eqnarray}\label{Tij wedge} 
\langle T^{i}_{\ j} \rangle_{\text{wedge}}=\frac{f(\Omega)}{r^d}\text{diag}\Big(1, -(d-1),1,..,1\Big).
\end{eqnarray}
This is indeed the case for free BCFTs \cite{Bissi:2022bgu, Deutsch:1978sc, Brevik:1998bya} and holographic BCFTs, as discussed in this paper. For simplicity, we have abbreviated $f(\theta, \Omega)$ as $f(\Omega)$ above. The function $f(\Omega)$ characterizes the properties of BCFTs and has the following interesting limits
\begin{eqnarray}\label{limits}
f(\Omega)\to \begin{cases}
\kappa_1/\Omega^d,\  \ \ \ \ \ \ \text{for } \Omega \to 0,\\
\kappa_2 (\pi-\Omega),\  \ \text{for } \Omega \to \pi.
\end{cases}
\end{eqnarray}
In the limits $r\to \infty, \Omega\to 0$ with $r \Omega=L$ fixed, the large $r$ region of a wedge approaches to a strip \cite{Diatlyk:2024zkk,Deutsch:1978sc}. In this way, we see that the constant  $\kappa_1$ is related to the Casimir effect of a flat strip with the width $L$
\begin{eqnarray}\label{Tij strip} 
\langle T^{i}_{\ j} \rangle_{\text{strip}}=\frac{\kappa_1}{L^d}\text{diag}\Big(1, -(d-1),1,..,1\Big). 
\end{eqnarray}
On the other hand, the physical meaning of $\kappa_2$ is unclear so far. This paper shows that  $\kappa_2$ is universally determined by the norm of the displacement operator. The displacement operator $D(y)$ describes the violation of the conservation of energy-stress tensors due to the boundary \cite{Billo:2016cpy}
\begin{eqnarray}\label{DT}
\nabla_i T^{ij}=-\delta(x)D^j(y),
\end{eqnarray}
where $D^j(y)=-D(y) n^j$ with $n^j$ is the outward-pointing normal vector on the boundary $x=0$. The Zamolodchikov norm of displacement operator is defined as \cite{Billo:2016cpy}
\begin{eqnarray}\label{Zamolodchikov norm} 
\langle D(y) D(0) \rangle =\frac{C_D}{|y|^{2d}}. 
\end{eqnarray}
We prove the following universal relation between the wedge Casimir effect and displacement operator 
\begin{eqnarray}\label{universal relation} 
\kappa_2= \frac{d\pi^{\frac{d-2}{2}}\Gamma(\frac{d}{2})}{2(d-1) \Gamma[d+2]} C_D,
\end{eqnarray}
and verify it with free BCFTs and AdS/BCFT. Besides, we argue that the Casimir energy density of the wedge $T_{tt}=-f(\Omega)/r^d$ increases with the opening angle 
\begin{eqnarray}\label{Monotonicity}
f'(\Omega)\le 0,
\end{eqnarray}
and test it with several examples. 

On the other hand, we investigate the holographic dual of the wedge in AdS/BCFT  \cite{Takayanagi:2011zk} \footnote{Note that we focus on the AdS/BCFT correspondence for wedges, where the wedge is on the AdS boundary. Please do not confuse it with wedge holography \cite{Akal:2020wfl, Miao:2020oey}, where the wedge is in bulk.}. See \cite{Akal:2020wfl, Miao:2020oey, Miao:2021ual, Cui:2023gtf, Ogawa:2022fhy, Miao:2022oas, Miao:2023mui, Suzuki:2024cqy, Aguilar-Gutierrez:2023tic, Aguilar-Gutierrez:2023zoi} for some recent developments of AdS/BCFT and its generalizations. Constructing the holographic dual of a BCFT with a fixed boundary shape in flat space is a non-trivial problem \footnote{We thank C. S. Chu for valuable discussions on this and related problems.}. Many typical solutions in AdS/CFT are not allowed in AdS/BCFT with a specific BCFT boundary. Let us explain the reasons. We impose the Neumann boundary condition (NBC) on the EOW brane $Q$  \cite{Takayanagi:2011zk} 
\begin{eqnarray}\label{NBC}
\Big(K^{ij}-(K-T)h^{ij} \Big)|_Q=0,
\end{eqnarray}
and Dirichlet boundary condition on the AdS boundary $M$
\begin{eqnarray}\label{DBC}
\delta g_{ij}|_M=0,
\end{eqnarray}
were $K$ is the extrinsic curvature and $T=(d-1) \tanh(\rho_*)$ is the brane tension 
\footnote{Following \cite{Takayanagi:2011zk}, we parameterize the brane tension as $T=(d-1) \tanh(\rho_*)$ in this paper. Here $\rho_*$ denotes the location of EOW brane $\rho=\rho_*$ in the metric $ds^2=d\rho^2 + \cosh^2(\rho)ds^2_{\text{AdS}_d}$  \cite{Takayanagi:2011zk}. We also set the AdS radius $l=1$ for simplicity. }.  
Note that this paper focuses on Takayanagi's model of AdS/BCFT \cite{Takayanagi:2011zk} with NBC (\ref{NBC}) on the brane. 
See \cite{Miao:2018qkc, Chu:2021mvq, Miao:2017gyt, Chu:2017aab} for discussions on other boundary conditions. Given a bulk metric obeying Einstein gravity in bulk $N$ and DBC (\ref{DBC}) on the AdS boundary $M$, we can fix the embedding function of EOW brane from the trace of NBC (\ref{NBC}), i.e., $(1-d)K+d T=0$. That is because the EOW brane is a codim-1 surface; we need only one equation to fix its location. For the less-symmetric case, the traceless parts of NBC $K_{ij}-(K/d)h_{ij}=0$ impose additional constraints apart from the trace. Not all given bulk solutions automatically satisfy these additional constraints. Thus, not all solutions in AdS/CFT are solutions in AdS/BCFT with a given BCFT geometry.

So far, we only know the gravity duals of BCFT in flat space with some symmetrical boundaries, such as the half-space, strip, and Euclidean disk \cite{Takayanagi:2011zk,Fujita:2011fp}. There are some perturbative discussions for the less-symmetrical boundaries with non-zero extrinsic curvatures \cite{Miao:2017aba,Miao:2018dvm}. In this paper, we construct the holographic dual of wedge space (\ref{metric}), providing an exact example of AdS/BCFT with a less-symmetric boundary. We remark that the Poincar\'e AdS cannot be dual to the wedge space. One can directly check that  Poincar\'e AdS disobeys the NBC (\ref{NBC}) for the BCFT living in a flat wedge space. See Appendix B. Another way to see this is that  Poincar\'e AdS has zero holographic stress tensors, which cannot produce the non-zero Casimir effect (\ref{Tij wedge0a},\ref{Tij wedge0b},\ref{Tij wedge}) of a wedge. One may guess Poincar\'e AdS could be dual to an excited state instead of a ground state of the wedge. This is not the case. The geometric singularity at the wedge corner is expected to produce a divergent expectation value of stress tensors near the corner generally, i.e., $\langle T^i_{\ j} \rangle \sim 1/r^d$ \footnote {It is similar to the case near a curved smooth boundary \cite{Deutsch:1978sc, Miao:2017aba}, i.e., $\langle T^i_{\ j} \rangle \sim \bar{k}^i_{\ j}/r^{d-1}$, where $\bar{k}^i_{\ j}$ are the traceless-parts of extrinsic curvatures. The divergent terms of $\langle T^i_{\ j} \rangle$ are universally determined by the Weyl anomaly and are independent of the states of BCFTs \cite{Miao:2017aba}.}. See section 2 for an example, where we prove the renormalized stress tensor takes a universal non-zero expression (\ref{Tij wedge universal}) in the limit $r\to 0, \Omega\to \pi$ for general cases. Thus, Poincar\'e AdS with zero Brown-York stress tensors cannot be dual to a wedge with non-zero $\langle T^i_{\ j} \rangle $. Note that we focus on the initial model of AdS/BCFT \cite{Takayanagi:2011zk} instead of the one \cite{Miyaji:2022dna, Biswas:2022xfw} with defects on the EOW branes. For the later model, the EOW brane is singular in bulk. As a result, the Poincar\'e AdS could be dual to the wedge for some parameters \cite{Miyaji:2022dna}. However, the above discussions imply these parameters are unphysical and should be ruled out. 

Let us summarize our holographic results. We construct the holographic dual of wedges in general dimensions. Our construction produces the expected wedge Casimir effect (\ref{Tij wedge}) together with the correct singular and smooth limits (\ref{limits}). In particular, it obeys the universal relation (\ref{universal relation}) between the displacement operator and  Casimir effects. These are solid supports for our results. Next, we consider the wedge contribution to the holographic entanglement entropy and find it increases with the opening angle, similar to the wedge Casimir energy. Finally, we discuss the holographic dual of polygons in AdS$_3$/BCFT$_2$ and its possible generalizations to higher dimensions.

The paper is organized as follows. 
In section 2, we study the universal feature of the Casimir effect of wedges and its relation to the displacement operator. In section 3, we formula the holographic dual of wedges and verify that it satisfies the universal relation (\ref{universal relation}). Section 4 investigates the wedge contributions to holographic entanglement entropy. Section 5 discusses the holographic duals of polygons in AdS$_3$/BCFT$_2$. Finally, we conclude with some open problems in section 6.

\section{Universal Casimir effect of wedges}

This section proves the universal relation (\ref{universal relation}) between the Casimir effect and the displacement operator and tests it by free BCFTs. Besides, we argue that $f'(\Omega)<0$ and verify it with several examples. This section mainly focuses on the field theory, and we will discuss the gravity dual in the next section. 

\subsection{Displacement operator}

Let us give a quick review of the displacement operator. Variating the effective action of BCFT with respect to the metric and the position of the boundary, we have \cite{Billo:2016cpy, McAvity:1993ue}
\begin{eqnarray}\label{dID}
\delta I_{\text{eff}}=\frac{1}{2}\int_M d^dx \sqrt{|g_M|} T^{ij}\delta g_{M\ ij}+\frac{1}{2}\int_{\partial M} d^{d-1}y \sqrt{|\sigma|}\left( t^{ij}\delta \sigma_{ij}+2D \delta x \right)
\end{eqnarray}
where $T^{ij}$, $t^{ij}$ are bulk and boundary stress tensors of BCFTs, respectively, and $D=-D^i n_i$ is the displacement operator. For simplicity, we focus on the flat Euclidean half space  
\begin{eqnarray}\label{flat half space}
ds^2=dx^2+\sum_{a=1}^{d-1} dy^a dy^a, \ \ x\ge 0,
\end{eqnarray}
where we turn off the variations of derivatives of the boundary metric $\partial_x^n h_{ij}$, such as the extrinsic curvatures. 

Consider the diffeomorphism in bulk
\begin{eqnarray}\label{diffbulk}
\delta_{\xi} y^a=0, \ \ \delta_{\xi} x^i=-\xi^i,\ \  \delta_{\xi} g_{M\ ij}=2\nabla_{(i} \xi_{j)},
\end{eqnarray}
we have 
\begin{eqnarray}\label{dIDbulk1}
\delta_{\xi}  I_{\text{eff}}&=&-\int_{ M} d^{d}x \sqrt{|g_M|} \xi_j \nabla_i T^{ij}\nonumber\\
&&+\int_{\partial M} d^{d-1}y \sqrt{|\sigma|} \left(T^{ij}n_i\xi_j -\hat{\xi}_j\hat{\nabla}_i t^{ij} +\xi_j n^j D \right)\nonumber\\
&=&0,
\end{eqnarray}
where $\hat{\nabla}_i$ is the induced covariant derivative on the boundary and $\hat{\xi}_i=\xi_l h^l_i$ is the pull back of the bulk vector $\xi_i$ into the boundary. We have used $t^{ij}n_j=0$ and the vanishing of extrinsic curvatures in the above derivations. For an infinite-small $\xi$ in bulk, we derive from (\ref{dIDbulk1})
\begin{eqnarray}\label{bulklaw}
\nabla_i T^{ij}=0, \ \text{for} \ x>0. 
\end{eqnarray}
As for the infinite-small $\xi$ on the boundary, we get \cite{Jensen:2015swa}
\begin{eqnarray}\label{boundarylaw1}
&&\hat{\nabla}_i t^{ij}=T^{il}n_i h_l^j , \\
&& D=-T^{ij}n_in_j. \label{boundarylaw2}
\end{eqnarray} 
Thus, the displacement operator $D=-T^{xx}|_{x=0}$ is given by the normal-normal component of the stress tensor in half plane  \cite{Billo:2016cpy, McAvity:1993ue, Jensen:2015swa}. By using this fact and the conformal symmetries, one can derive the one-point function $\langle D(y) \rangle=0$ and two-point functions  \cite{Billo:2016cpy, McAvity:1993ue} in flat half space,
\begin{eqnarray}\label{TD1}
&&\langle T^{xx}(x_1) D(y)\rangle= -\frac{d C_D}{(d-1)}\ \frac{1}{(x_1^2+y^2)^{d}}\left( \frac{(x_1^2-y^2)^2}{(x_1^2+y^2)^{2}}-\frac{1}{d}\right), \\ \label{TD2}
&&\langle T^{ab}(x_1) D(y)\rangle=-\frac{d C_D}{(d-1)}\ \left( \frac{4x_1^2 y^ay^b}{(x_1^2+y^2)^{d+2}}-\frac{\delta^{ab}}{d(x_1^2+y^2)^{d}}\right),\\ \label{TD3}
&&\langle T^{ax}(x_1) D(y)\rangle=-\frac{2d C_D}{(d-1)}\ \left( \frac{ y^a x_1}{(x_1^2+y^2)^{d+1}}-\frac{2y^{a}x_1^3}{(x_1^2+y^2)^{d+2}}\right),
\end{eqnarray} 
where $T^{ij}(x_1)=T^{ij}(x_1,y_1^a=0)$ and $D(y)=D(x=0,y^a)$.  Note that, in the limit $x_1\to 0$,  (\ref{TD1}) reduces to the two-point function of displacement operator (\ref{Zamolodchikov norm}). 

\subsection{Universality in smooth limit}

According to \cite{Bianchi:2016xvf}, one can derive the one-point function of an operator near the deformed boundary from the two-point function of this operator with displacement operator on the non-deformed boundary. For our case of the stress tensor, we have
\begin{eqnarray}\label{keyformulaD}
\langle T_{ij}(x_1)\rangle_{f}=\langle T_{ij}(x_1)\rangle_{0}-\int d^{d-1}y\langle T_{ij}(x_1)D(y)\rangle f(y)+O(f^2),
\end{eqnarray}
where $\delta x=f(y)$ denotes the small deformation of the boundary, $\langle T_{ij}(x_1)\rangle_{0}=0$ in the half space with undeformed boundary, and we have set $y_1^a=0$ for $T_{ij}$ and $x=0$ for $D$ for simplicity.  

We are ready to derive the Casimir effect of wedges in the smooth limit $\Omega\to \pi$. By slightly deforming the plane boundary
\begin{eqnarray}\label{deform function}
\delta x=f(y)=\Pi(y^1) y^1 (\Omega-\pi),
\end{eqnarray}
we get a wedge with opening angle $\Omega\to \pi$. Here $\Pi(y^1)$ is the step function, equal to one when $y^1\ge 0$ and zero otherwise. Substituting (\ref{deform function}) and correlation functions (\ref{TD1},\ref{TD2},\ref{TD3}) into (\ref{keyformulaD}) and performing the coordinate transformations $\rho^2=\sum_{a=2}^d y^ay^a$, we derive
\begin{eqnarray}\label{Txx}
\langle T_{x x}(x_1)\rangle_{f}&=&\frac{2 \pi ^{\frac{d-2}{2}}}{\Gamma \left(\frac{d-2}{2}\right)} \int_0^{\infty} \rho^{d-3} d\rho \int_0^{\infty} dy^1 y^1 (\Omega-\pi)\nonumber\\
&& \times\frac{d C_D}{(d-1)}\ \frac{1}{(x_1^2+\rho^2+{y^1}^2)^{d}}\left( \frac{(x_1^2-\rho^2-{y^1}^2)^2}{(x_1^2+\rho^2+{y^1}^2)^{2}}-\frac{1}{d}\right)\nonumber\\
&=&\frac{d \pi ^{\frac{d-2}{2}} \Gamma \left(\frac{d}{2}\right)C_D }{2 (d-1) \Gamma (d+2)} \frac{(\pi-\Omega) }{x_1^d}+O(\pi-\Omega)^2,
\end{eqnarray}
\begin{eqnarray}\label{Tyy}
\langle T_{y^1 y^1}(x_1)\rangle_{f}&=&\frac{2 \pi ^{\frac{d-2}{2}}}{\Gamma \left(\frac{d-2}{2}\right)} \int_0^{\infty} \rho^{d-3} d\rho \int_0^{\infty} dy^1 y^1 (\Omega-\pi)\nonumber\\
&& \times \frac{d C_D}{(d-1)}\ \left( \frac{4x_1^2 y^1y^1}{(x_1^2+\rho^2+{y^1}^2)^{d+2}}-\frac{1}{d(x_1^2+\rho^2+{y^1}^2)^{d}}\right)\nonumber\\
&=&-(d-1) \frac{d \pi ^{\frac{d-2}{2}} \Gamma \left(\frac{d}{2}\right)C_D}{2 (d-1) \Gamma (d+2)}  \frac{(\pi-\Omega) }{x_1^d}+O(\pi-\Omega)^2,
\end{eqnarray}
and for $a\ne 1$
\begin{eqnarray}\label{Tab}
\langle T_{y^a y^a}(x_1)\rangle_{f}&=&\frac{2 \pi ^{\frac{d-2}{2}}}{\Gamma \left(\frac{d-2}{2}\right)} \int_0^{\infty} \rho^{d-3} d\rho \int_0^{\infty} dy^1 y^1 (\Omega-\pi)\nonumber\\
&& \times \frac{d C_D}{(d-1)}\ \left( \frac{4x_1^2 \rho^2/(d-2)}{(x_1^2+\rho^2+{y^1}^2)^{d+2}}-\frac{1}{d(x_1^2+\rho^2+{y^1}^2)^{d}}\right)\nonumber\\
&=&\frac{d \pi ^{\frac{d-2}{2}} \Gamma \left(\frac{d}{2}\right) C_D}{2 (d-1) \Gamma (d+2)}  \frac{(\pi-\Omega) }{x_1^d}+O(\pi-\Omega)^2.
\end{eqnarray}
Similarly, one can check that the off-diagonal elements of stress tensors vanish, i.e., $\langle T_{x y^a}(x_1)\rangle_{f}=\langle T_{y^a y^b}(x_1)\rangle_{f}=0$. The above results can be rewritten into a unite form 
\begin{eqnarray}\label{Tij form D}
\langle T^i_j (x_1)\rangle_{f}= \frac{d \pi ^{\frac{d-2}{2}} \Gamma \left(\frac{d}{2}\right)C_D }{2 (d-1) \Gamma (d+2)} \frac{(\pi-\Omega) }{x_1^d} \text{diag}\Big(1, -(d-1),1,...,1 \Big)+O(\pi-\Omega)^2. 
\end{eqnarray}
Comparing (\ref{Tij form D}) with (\ref{Tij wedge}, \ref{limits}) and identifying $x_1$ and $r$ in the smooth limit $\Omega\to \pi$, we finally obtain the universal relation (\ref{universal relation}). 

Some comments are in order. 
\begin{itemize}
  \item The above discussions focus on flat space with a slightly deformed boundary. It can be generalized to curved wedge space as long as we focus on the region near the boundary, i.e., $r\ll 1/\sqrt{R}$, where $r$ is the distance to the boundary and $R$ labels the curvature of curved wedge space. That is because the spacetime is locally flat, and the two-point functions (\ref{TD1},\ref{TD2},\ref{TD3}) still work in the near-boundary region for local field theories. Our results imply a universal Casimir effect near the wedge $r\sim 0$ 
\begin{eqnarray}\label{Tij wedge universal} 
\langle T^{i}_{\ j} \rangle_{\text{wedge}}=\frac{d\pi^{\frac{d-2}{2}}\Gamma(\frac{d}{2})C_D}{2(d-1) \Gamma[d+2]} \frac{(\pi-\Omega)}{r^d}\text{diag}\Big(1, -(d-1),1,..,1\Big)+O\Big(\frac{1}{r^{d-1}}, (\pi-\Omega)^2\Big), 
\end{eqnarray}
in the smooth limit $\Omega\sim \pi$ in general curved wedge space, where $r$ is proper distance to the corner of the wedge. 

 \item  The above result (\ref{Tij wedge universal}) applies to a singular boundary, which reminds the universal Casimir effect of BCFT near a smooth boundary \cite{Miao:2017aba, Miao:2018dvm}
\begin{eqnarray}\label{Tij from Weyl anomaly}
\langle T_{ij} \rangle  = \alpha \frac{\bar{k}_{ij}}{r^{d-1}}+..., \ \ r\sim 0,
\end{eqnarray}
where $\bar{k}_{ij}$ are the traceless parts of extrinsic curvatures, $r$ is the proper distance to the boundary, and $\alpha $ is given by the norm of displacement operator \cite{Miao:2018dvm}
\begin{eqnarray}\label{alpha CD}
\alpha=-\frac{d\Gamma[\frac{d+1}{2}]\pi^{\frac{d-1}{2}}}{(d-1)\Gamma[d+2]}C_D. 
\end{eqnarray}
Interestingly, (\ref{Tij wedge universal}) and (\ref{Tij from Weyl anomaly}) depend on both $C_D$, which is related to the B-type boundary central charge of Weyl anomaly \cite{Miao:2017aba, Miao:2018dvm, Herzog:2017kkj, Herzog:2017xha}. Since Weyl anomaly is universal and is independent of the states of BCFTs \footnote{The vacuum state, thermal state, and excited states of BCFTs all have the same Weyl anomaly in a fixed background spacetime.}, so does the universal Casimir effects (\ref{Tij wedge universal}) and (\ref{Tij from Weyl anomaly}). 

 \item  Let us test the universal relation (\ref{universal relation}) by free BCFTs. According to \cite{Deutsch:1978sc}, we have for 4d free BCFTs 
 \begin{eqnarray}\label{wedge Casimir free BCFTs}
f(\Omega)=\begin{cases}
\frac{1}{1440 \Omega^2}(\frac{\pi^2}{\Omega^2}-\frac{\Omega^2}{\pi^2}),\  \ \ \ \ \ \ \ \ \  \ \text{free scalar},\\
\frac{1}{720 \pi^2}(\frac{\pi^2}{\Omega^2}+11)(\frac{\pi^2}{\Omega^2}-1),\  \ \text{Maxwell theory},
\end{cases}
\end{eqnarray}
which yields
\begin{eqnarray}\label{kappa2 free BCFTs}
\kappa_2=\begin{cases}
\frac{1}{360 \pi^3},\  \ \ \ \ \ \ \text{free scalar},\\
\frac{1}{30 \pi^3},\  \ \ \ \ \ \ \ \text{Maxwell theory}.
\end{cases}
\end{eqnarray}
It's worth noting that (\ref{wedge Casimir free BCFTs}) holds for both Dirichlet and Robin(Neumann) boundary conditions (BC) for free scalar and applies to both Absolute and Relative BCs for Maxwell theory. Again, we stress that we impose the same BCs on the two boundaries of the wedge. If we choose different BCs on the two wedge boundaries, the discontinuous BCs induce a new singularity in addition to the geometric singularity at the corner of the wedge. We leave this non-trivial case to future work. 
The corresponding $C_D$ of free BCFTs can be read off from \cite{Miao:2017aba,Herzog:2017kkj,Herzog:2017xha,Fursaev:2015wpa,Herzog:2015ioa}. We have
\begin{eqnarray}\label{CD free BCFTs}
C_D=\begin{cases}
\frac{1}{2 \pi ^4},\  \ \ \ \ \ \ \text{free scalar},\\
\frac{6}{\pi ^4},\  \ \ \ \ \ \ \ \text{Maxwell theory}.
\end{cases}
\end{eqnarray}
The above two equations give $\kappa_2/C_D=\pi/180$, which agrees with the universal relation (\ref{universal relation}) for $d=4$. $C_D$ for conformally coupled scalar in general dimensions is given by \cite{Miao:2018dvm}
\begin{eqnarray}\label{CD higher D scalar}
C_D=\frac{\Gamma[\frac{d}{2}]^2}{2\pi^d}.
\end{eqnarray}
The corresponding $\kappa_2$ can be read off from (\ref{Mirror: f even d 1},\ref{Mirror: f 3d smooth limit},\ref{Mirror: f 5d result},\ref{Mirror: f 7d result})
 of appendix A. We list $\kappa_2$ and $C_D$ in Table. \ref{table1kappa2}, and verify that the universal relation (\ref{universal relation}) is obeyed in higher dimensions too. 
 
  \begin{table}[ht]
\caption{$\kappa_2$ and $C_D$ for conformally coupled scalar}
\begin{center}
    \begin{tabular}{| c | c | c | c |  c | c | c | c| c| c|c| }
    \hline
     d &$2$&3& $4$ & $5$ & 6 &7 &8  \\ \hline
  $\kappa_2$ & $\frac{1}{12 \pi ^2}$ & $\frac{1}{512 \pi}$& $\frac{1}{360 \pi ^3}$ & $\frac{3}{16384\pi^2}$ & $\frac{1}{2100 \pi ^4}$ & $\frac{25}{524288\pi^3}$ & $\frac{1}{5880 \pi ^5}$ \\ \hline
  $C_D$ & $\frac{1}{2 \pi ^2}$ & $\frac{1}{8 \pi ^2}$& $ \frac{1}{2 \pi ^4}$ & $ \frac{9}{32 \pi ^4}$ & $\frac{2}{\pi ^6}$ & $\frac{225}{128 \pi ^6} $ & $\frac{18}{\pi ^8}$ \\ \hline
     \end{tabular}
\end{center}
\label{table1kappa2}
\end{table}
 
  \item  As we will show in the next section, the universal relation (\ref{universal relation}) also holds for AdS/BCFT.  
 
 \end{itemize}
 
\subsection{Monotonicity of wedge Casimir effect}

Now we provide a simple argument that $f'(\Omega)\le  0$. Without loss of generality, we focus on the range $0\le \Omega\le \pi$ with $f(\Omega)\ge 0$ below. The range $\pi\le \Omega\le 2\pi$ with $f(\Omega)\le 0$ is similar.  In the following argument, the key point is that the energy density $T_{tt}$ and the pressure $T_{\theta\theta}$ have the same sign. On the other hand, the sign of $T_{tt}\sim -f(\Omega)$ is unimportant. 

Suppose we adiabatically make larger the wedge by $d\Omega>0$. Since the pressure along the $\theta$ direction is negative, we need to do positive work to enlarge the wedge
\begin{eqnarray}\label{monotonicity: pdV}
dW=-p dV \sim -T_{\theta\theta} d\Omega\sim (d-1) \frac{f(\Omega)}{r^d} d\Omega \ge 0.
\end{eqnarray}
As a result, the total energy of the wedge should become larger. However,  since the energy density is negative
\begin{eqnarray}\label{monotonicity: E}
T_{tt}=-\frac{f(\Omega)}{r^d} \le 0, 
\end{eqnarray}
the total energy $\int T_{tt} dV$ becomes smaller if the energy density remains invariant. To have a larger energy, the energy density $T_{tt}$ must becomes larger, which yields
\begin{eqnarray}\label{monotonicity: dE}
f'(\Omega)\le 0. 
\end{eqnarray}

Below, let us test the inequality (\ref{monotonicity: dE}). From (\ref{limits}), we have in the singular and smooth limits
\begin{eqnarray}\label{monotonicity: limits}
f'(\Omega)\to \begin{cases}
-\frac{d \kappa_1}{\Omega^{d+1}},\  \ \ \text{for } \Omega \to 0,\\
-\kappa_2 ,\  \ \ \ \ \ \text{for } \Omega \to \pi,
\end{cases} < 0,
\end{eqnarray}
which indeed obeys (\ref{monotonicity: dE}).  Above, we have used $\kappa_1\ge 0$ that the Casimir energy of parallel plates is negative, and $\kappa_2\sim C_D \ge 0$ that the norm of the displacement operator is positive. The inequality (\ref{monotonicity: dE}) also holds for free 4d BCFTs
\begin{eqnarray}\label{monotonicity: free BCFTs}
f'(\Omega)=\begin{cases}
-\frac{\pi ^2}{360 \Omega ^5},\  \ \ \ \ \ \ \ \ \  \ \text{free scalar},\\
-\frac{5 \Omega ^2+\pi ^2}{180 \Omega ^5},\  \ \text{Maxwell theory},
\end{cases} <0,
\end{eqnarray}
and holographic BCFT with zero brane tension (\ref{zero tension: f Omega})
\begin{eqnarray}\label{monotonicity: AdSBCFT}
f'(\Omega)=-\frac{2 \pi ^2 \Omega ^{d-3} \left(\frac{\sqrt{(d-2) d \Omega ^2+\pi ^2}-\pi }{d-2}\right)^{1-d}}{\sqrt{(d-2) d \Omega ^2+\pi ^2}} <0.
\end{eqnarray}
Fig. \ref{3d4dwedgeCasimir} suggests that the inequality (\ref{monotonicity: dE}) also works for the AdS/BCFT with non-zero brane tension. 

In summary, in the smooth limit, we prove the universal relation (\ref{universal relation}) between the displacement operator and the wedge Casimir effect. Besides, we argue that the wedge Casimir energy density increases with the opening angle, i.e., $f'(\Omega) \le 0$. Finally, we test our results with free BCFTs. We will also test them using AdS/BCFT in the next section.

\section{Gravity dual of wedges}

In this section, we construct the holographic dual of wedges in AdS/BCFT. 
We find it is given by a cutout of the AdS soliton. 
We stress again that we focus on AdS/BCFT \cite{Takayanagi:2011zk} instead of wedge holography  \cite{Akal:2020wfl, Miao:2020oey}
 in this paper. We verify that our construction produces the expected Casimir effect of wedges (\ref{Tij wedge}, \ref{limits}). In particular, it obeys the universal relation (\ref{universal relation}) between the Casimir effect and the displacement operator. 

\subsection{General theory}

We aim to find the holographic dual of a BCFT living in the wedge space (\ref{metric}).  As we have discussed in the introduction, it cannot be the Poincar\'e AdS since it does not obey NBC (\ref{NBC}) and cannot produce the correct Casimir effect (\ref{Tij wedge}).  Instead, we find 
the AdS soliton 
can do the job
\begin{eqnarray}\label{dual: metric}
ds^2=\frac{\frac{dz^2}{h(z)}+h(z) d\theta^2+\frac{dr^2+\sum_{\hat{i},\hat{j}=1}^{d-2} \eta_{\hat{i}\hat{j}} dy^{\hat{i}} dy^{\hat{j}}}{r^2}}{z^2},
\end{eqnarray}
where $h(z)=1-z^2-c_1 z^d$. We label the constant $c_1=(1-z_h^2)/z_h^d$ so that $h(z_h)=0.$ To be absence of conical singularity in bulk,  we choose the angle period in bulk as
\begin{eqnarray}\label{dual: period}
\beta=\frac{4\pi}{|h'(z_h)|}=\frac{4\pi z_h}{d+(2-d) z_h^2}.
\end{eqnarray}
On the other hand, the range of angle is $0\le \theta\le \Omega < \beta$ on the AdS boundary $z\to 0$. 
It should be mentioned that the metric takes the form of a hyperbolic black hole (with time replaced by angle $\theta$), which plays a vital role in holographic R\'enyi entropy \cite{Hung:2011nu, Dong:2016fnf} and cone holography  \cite{Miao:2021ual, Cui:2023gtf}.  

Let us show the metric (\ref{dual: metric}) has the expected features to be dual to a wedge.  On the AdS boundary $z\to 0$, the BCFT metric is conformally equivalent to the wedge metric (\ref{metric}) up to a Weyl factor $1/r^2$. To get exactly the wedge metric (\ref{metric}), One can perform a Weyl transformation on the AdS boundary, equivalently, a coordinate transformation in bulk.  The coordinate transformations for $d=2$ are
\begin{eqnarray}\label{dual: coordinate transformation 2d z}
&&z=\frac{4 r_1 z_1}{\sqrt{8 \left(c_1+2\right) r_1^2 z_1^2+c_1^2 z_1^4+16 r_1^4}}, \\
&&r=r_1 \left(\frac{\left(2 \sqrt{c_1+1}+c_1+2\right) z_1^2+4 r_1^2}{\left(-2 \sqrt{c_1+1}+c_1+2\right) z_1^2+4 r_1^2}\right){}^{\frac{1}{2 \sqrt{c_1+1}}}, \label{dual: coordinate transformation 2d r}
\end{eqnarray}
which yields the bulk metric
\begin{eqnarray}\label{dual: metric 2d}
ds^2=\frac{dz_1^2+\Big(1+f_r(\frac{z_1}{r_1})\Big) dr_1^2+r_1^2\Big(1+f_{\theta}(\frac{z_1}{r_1})\Big)d\theta^2}{z_1^2},
\end{eqnarray}
with $f_r(z_1)=\frac{1}{16} c_1 z_1^2 \left(c_1 z_1^2+8\right)$ and $f_{\theta}(z_1)=\frac{1}{16} c_1 z_1^2 \left(c_1 z_1^2-8\right)$.  Clearly, the above metric gives the expected wedge metric $ds^2=dr_1^2+r_1^2 d\theta^2$ and BCFT stress tensor on the AdS boundary
\begin{eqnarray}\label{dual: Tij wedge 2d} 
\langle T^{i}_{\ j} \rangle_{d=2}=\frac{c_1}{r_1^2}\text{diag}\Big(1, -1\Big). 
\end{eqnarray}
For $d>2$, it is difficult to find the exact coordinate transformations. Instead, we obtain the perturbative ones up to $O(z_1^{d+1})$
\begin{eqnarray}\label{dual: coordinate transformation d z}
&&z=\frac{z_1}{\sqrt{r_1^2+z_1^2}}\Big(1-\frac{c_1 z_1^d}{2 d r_1^d}+O(z_1^{d+1}) \Big),\nonumber\\
&&r=\sqrt{r_1^2+z_1^2}\Big(1+O(z_1^{d+1}) \Big), \label{dual: coordinate transformation d r}
\end{eqnarray}
which transforms the bulk metric to be
\begin{eqnarray}\label{dual: metric d}
ds^2=\frac{dz_1^2+\Big(1+\frac{c_1z_1^d}{d r_1^d}\Big) \Big(dr_1^2+\sum_{\hat{i},\hat{j}=1}^{d-2} \eta_{\hat{i}\hat{j}} dy^{\hat{i}} dy^{\hat{j}}\Big) +r_1^2\Big(1-\frac{c_1(d-1)z_1^d}{d r_1^d}\Big)d\theta^2+O(z_1^{d+1})}{z_1^2}.
\end{eqnarray}
The above metric gives the expected wedge metric (\ref{metric}) and renormalized stress tensor (\ref{Tij wedge}) with 
\begin{eqnarray}\label{dual: c1 f} 
f(\Omega)=c_1.
\end{eqnarray}
In the above discussions, we have used the  formula of holographic stress tensor (setting $16\pi G_N=1$)\cite{deHaro:2000vlm}
\begin{eqnarray}\label{dual: Tij holo formu} 
\langle T_{i j} \rangle=d g^{(d)}_{ij},
\end{eqnarray}
where $ g^{(d)}_{ij}$ is defined in the Fefferman-Graham expansion of the asymptotic AdS space
\begin{eqnarray}\label{dual: FG} 
ds^2=\frac{dz^2+ (g^{(0)}_{ij}+z^2g^{(2)}_{ij}+...+z^dg^{(d)}_{ij}+...)dy^i dy^j}{z^2}.
\end{eqnarray}
In general, there are corrections to the formula (\ref{dual: Tij holo formu}) from Weyl anomaly for even $d$. For our case of flat space with zero Weyl anomaly, these potential corrections disappear.  Below we mainly focus on the coordinates $(z, r)$ instead of $(z_1, r_1)$, since the corresponding metric (\ref{dual: metric} has an exact expression.

\begin{figure}[t]
\centering
\includegraphics[width=10cm]{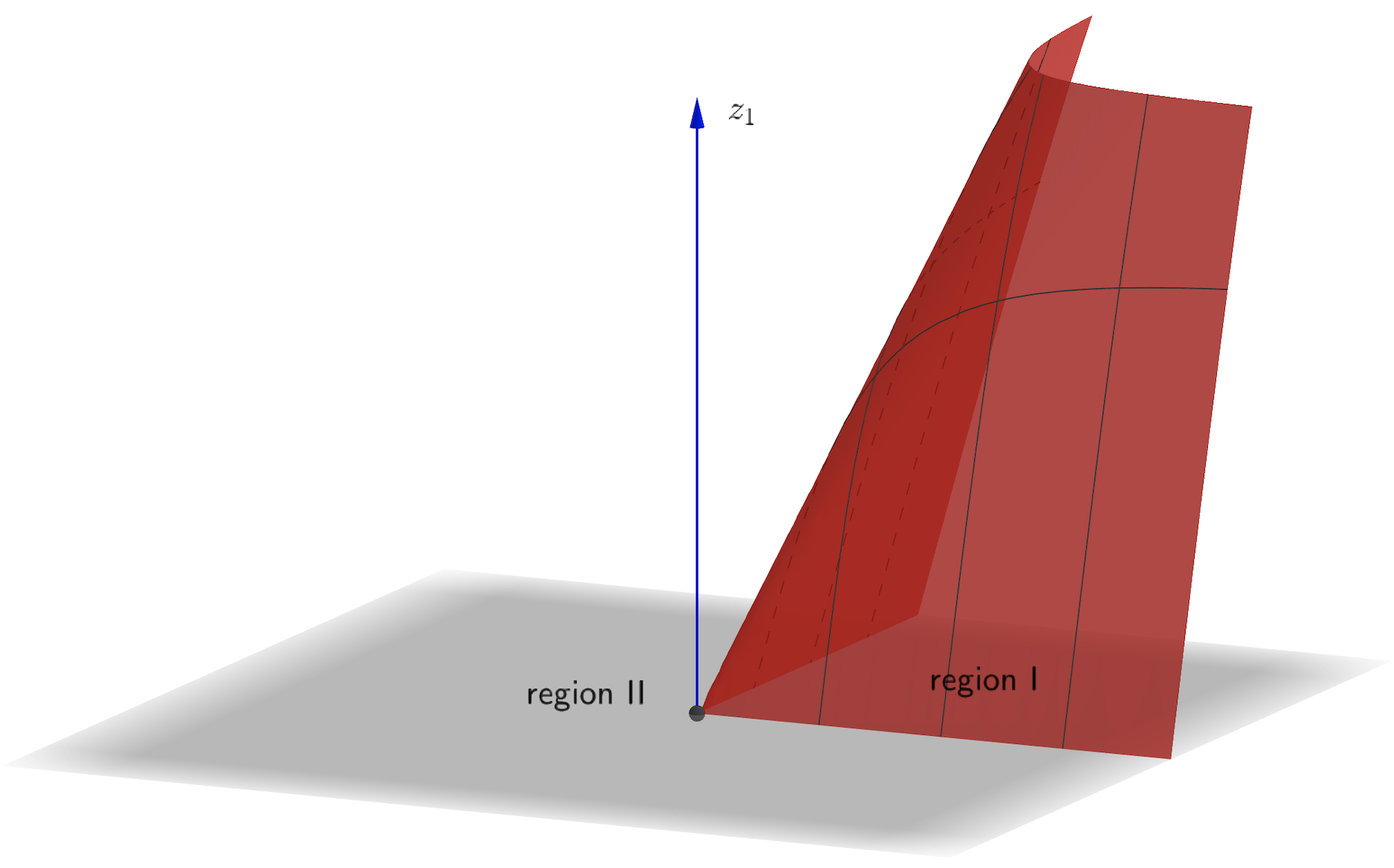}
\caption{Geometry for holographic dual of wedge in coordinates $(z_1, r_1)$. The gray plane denotes the AdS boundary $M$, which is divided by region I and region II.  BCFT can be defined either on region I or region II. Recall that the angle period is $\beta$ (\ref{dual: period}), which is not necessarily $2\pi$. Thus, the opening angles of region I and region II can be both smaller than $\pi$. Region I and region II share the same smooth EOW brane $Q$ (the red surface). Since extrinsic curvatures flip signs when crossing the red surface, from NBC (\ref{NBC}), the brane tension differs by a minus sign for the BCFT defined on region I and region II. Thus, once we obtain the AdS/BCFT of wedges with negative brane tension, by considering it's complement, we automatically get the one with positive brane tension. }
\label{wedgeAdSBCFT}
\end{figure}

Next, we show the metric (\ref{dual: metric}) obeys NBC (\ref{NBC}) for suitable embedding function of EOW brane
\begin{eqnarray}\label{dual: Q} 
z=z(\theta).
\end{eqnarray}
 As we have stressed in the introduction, the NBC for a fixed BCFT geometry is not always satisfied by a given bulk metric such as Poincar\'e AdS.  See Fig. \ref{wedgeAdSBCFT} for the schematic geometry of AdS/BCFT for wedges.  BCFT can be defined either on region I or region II on the AdS boundary $M$. These two regions share the same EOW brane $Q$ (red surface). Since extrinsic curvatures change signs when crossing a surface, from NBC (\ref{NBC}), the brane tension differs by a minus sign for the BCFT defined on region I and region II. Thus, once we obtain the AdS/BCFT of wedges with negative tension, by considering it's complement, we automatically get the one with positive tension. It is similar to the case of AdS/BCFT for strips \cite{Fujita:2011fp}. Without loss of generality, we focus on the region I with negative brane tension below.  We remark that both the opening angles of region I and region II can be smaller than $\pi$, since the angle period $\beta$ (\ref{dual: period}) is not necessarily $2\pi$. In fact, as we will show below, the opening angles are the same for the case of $d=2$, i.e., $\Omega_{\text{I}}=\Omega_{\text{II}}=\pi/\sqrt{1+c_1}$.

\begin{figure}[t]
\centering
\includegraphics[width=7.5cm]{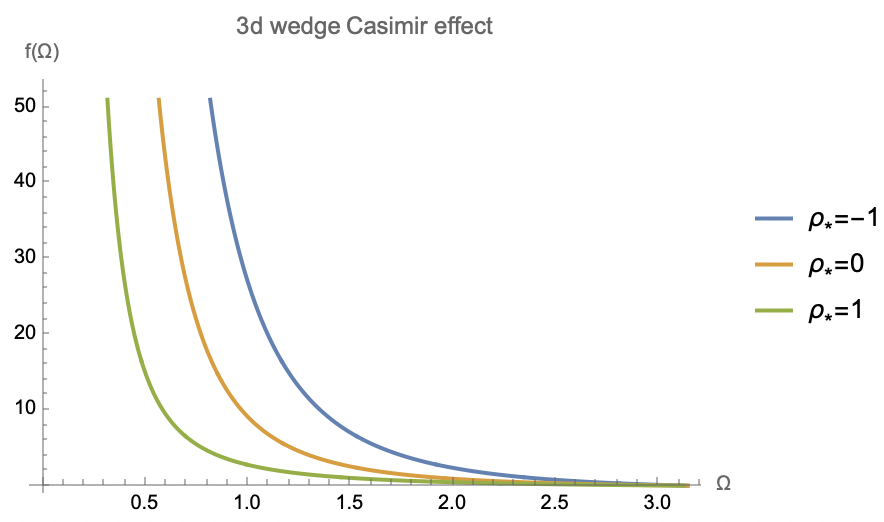} \includegraphics[width=7.5cm]{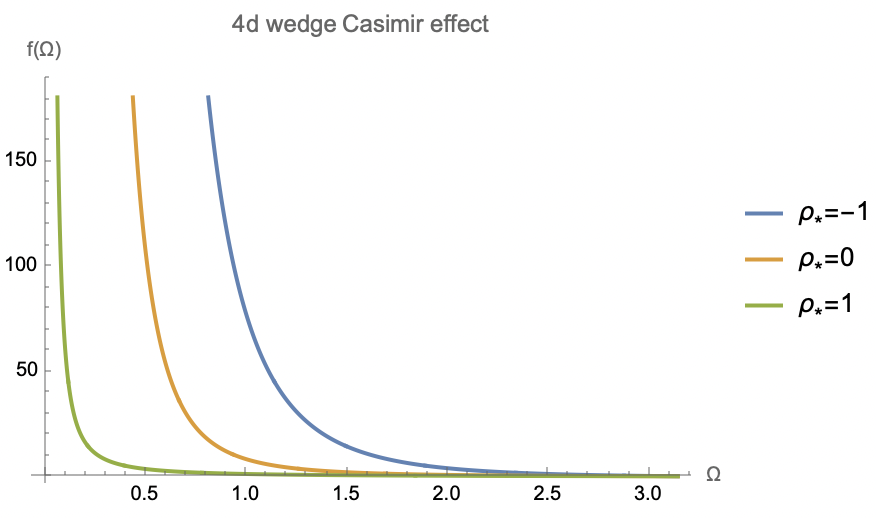}
\caption{Holographic Casimir effects of 3d wedge (left) and 4d wedge (right). Here, $\rho_*=-1, 0,1$ denote the negative, zero, and positive brane tensions. It shows $f(\Omega)$ approaches zero at $\Omega\to \pi$ and infinity at $\Omega \to 0$. Besides, it suggests $f(\Omega)$ decreases with the tension $T=(d-1) \tanh(\rho_*)$ and the opening angle $\Omega$.}
\label{3d4dwedgeCasimir}
\end{figure}

Substituting the embedding function (\ref{dual: Q}) and metric (\ref{dual: metric}) into NBC (\ref{NBC}), we get an independent equation for the region I with a negative brane tension $T=(d-1) \tanh(\rho_*)$
\begin{eqnarray}\label{dual: EOM Q} 
\frac{h(z(\theta ))}{\sqrt{\frac{z'(\theta )^2}{h(z(\theta ))}+h(z(\theta ))}}+\tanh \left(\rho _*\right)=0.
\end{eqnarray}
At the turning point $z(\Omega/2)=z_{\text{max}}$, we have $z'(\Omega/2)=0$. Then the above equation becomes 
\begin{eqnarray}\label{dual: EOM Q1} 
\sqrt{h(z_{\text{max}})}+\tanh \left(\rho _*\right)=0,
\end{eqnarray}
which yields
\begin{eqnarray}\label{dual: EOM Q1 c1} 
c_1=f(\Omega)=z_{\max }^{-d} \left(\text{sech}^2\left(\rho _*\right)-z_{\max }^2\right).
\end{eqnarray}
From (\ref{dual: EOM Q}) and (\ref{dual: EOM Q1}), we derive
\begin{eqnarray}\label{dual: theta} 
d\theta=\pm \frac{dz}{h(z) \sqrt{\frac{h(z)}{h\left(z_{\max }\right)}-1}},
\end{eqnarray}
which yields the embedding function
\begin{eqnarray}\label{dual: embedding function result} 
\theta_{<}(z)=\int_0^{z}\frac{ds}{h(s) \sqrt{\frac{h(s)}{h\left(z_{\max }\right)}-1}}, \ \ \text{for} \ 0\le \theta\le \Omega/2.
\end{eqnarray}
The case of $\Omega/2 \le \theta\le \Omega$ can be obtained by symmetry, i.e. $\theta_{>}(z)=\Omega-\theta_{<}(z)$. From (\ref{dual: embedding function result}), we obtain the opening angle for region I with negative $\rho_*$
\begin{eqnarray}\label{dual: opening angle result 1} 
\Omega_{\text{I}}(\rho_*)=2\theta(z_{\max})=2\int_0^{z_{\text{max}}}\frac{dz}{h(z) \sqrt{\frac{h(z)}{h\left(z_{\max }\right)}-1}}.
\end{eqnarray}
Recall that region II is the complement of region I. We get the opening angle for region II with positive $\rho_*$ \begin{eqnarray}\label{dual: opening angle result 2} 
\Omega_{\text{II}}(\rho_*)=\beta-\Omega_{\text{I}}(\rho_* \to -\rho_*),
\end{eqnarray}
where we have used the fact that the tensions of region I and region II differ by a minus sign.  From (\ref{dual: EOM Q1 c1}), we solve $z_{\text{max}}$ in terms of $f(\Omega)$ and then substitute it into (\ref{dual: opening angle result 1}).  Then, (\ref{dual: opening angle result 1}) becomes the inverse function of $f(\Omega)$, which can be obtained numerically generally. In fact, the integral expressions (\ref{dual: embedding function result},\ref{dual: opening angle result 1}) are good enough for many purposes. For instance, they can draw the figures of $f(\Omega)$.  See Fig. \ref{3d4dwedgeCasimir}, which shows that $f(\Omega)$ decreases with the tension $T=(d-1) \tanh(\rho_*)$ and the opening angle $\Omega$. 

\subsection{Exact examples}

Let us study some examples, where the integrals (\ref{dual: embedding function result},\ref{dual: opening angle result 1}) can be calculated exactly. 

{\bf Case I: } $d=2$

For $d=2$, we have $\beta=\frac{2 \pi }{\sqrt{c_1+1}}$ and $z_{\max }=\frac{\text{sech}\left(\rho _*\right)}{\sqrt{c_1+1}}$. Integrating (\ref{dual: theta},\ref{dual: opening angle result 1}, \ref{dual: opening angle result 2}), we obtain
\begin{eqnarray}\label{dual: 2d case angle}
&&\theta_{<}(z)=\frac{\tan ^{-1}\left(\frac{z}{\sqrt{\frac{\text{csch}^2\left(\rho _*\right)}{c_1+1}-z^2 \coth ^2\left(\rho _*\right)}}\right)}{\sqrt{c_1+1}},\\
&&\Omega_{\text{I}}=\Omega_{\text{II}}=\frac{\pi }{\sqrt{c_1+1}}=\frac{\beta}{2}. \label{dual: 2d case opening angle}
\end{eqnarray}
From (\ref{dual: 2d case opening angle}), we get the characteristic function of 2d wedge
\begin{eqnarray}\label{dual: f 2d}
f_{2d}(\Omega)=c_1=\frac{\pi ^2-\Omega ^2}{\Omega ^2},
\end{eqnarray}
which agrees with the results of field theory \cite{Cardy:1988tk}
\begin{eqnarray}\label{dual: f 2d BCFT}
f_{2d}(\Omega)=\frac{c}{24\pi} \frac{\pi ^2-\Omega ^2}{\Omega ^2},
\end{eqnarray}
where $c=\frac{3l}{2G_N}=24 \pi$ is the central charge of 2d BCFT. Recall that we have set $16\pi G_N=1$ and $l=1$. Since a wedge can be obtained by the conformal transformation of half space for $d=2$, the wedge Casimir effect takes universal form (\ref{dual: f 2d BCFT}) \cite{Cardy:1988tk}. This is indeed the case for the holographic result (\ref{dual: f 2d}).

{\bf Case II: } $d=4$

We first express the bulk angle period $\beta$ and turning point $z_{\max}$ in functions of $c_1$
\begin{eqnarray}\label{dual: 4d parameters}
\beta=\frac{2 \sqrt{2} \pi }{\sqrt{\frac{\left(4 c_1+1\right) \left(\sqrt{4 c_1+1}-1\right)}{c_1}}}, \ \ z_{\max }=\frac{\sqrt{\frac{\sqrt{4 c_1 \text{sech}^2\left(\rho _*\right)+1}-1}{c_1}}}{\sqrt{2}}. 
\end{eqnarray}
From (\ref{dual: embedding function result}), we obtain the embedding function for $0\le \theta\le \Omega/2$
\begin{eqnarray}\label{dual: 4d case angle}
\theta_{<}(z)&=&\frac{-z_{\max } \sqrt{\frac{c_1 \left(z_{\max }^4-z^4\right)+z_{\max }^2-z^2}{c_1 z_{\max }^4+z_{\max }^2}}}{2 \sqrt{\left(4 c_1+1\right) \left(\frac{c_1 z^4+z^2-1}{c_1 z_{\max }^4+z_{\max }^2-1}-1\right)}} \times \nonumber\\
&& \Big[(\sqrt{4 c_1+1}-1) \Pi \left(\frac{-2 c_1 z_{\max }^2}{\sqrt{4 c_1+1}+1};-\sin ^{-1}\left(\frac{z}{z_{\max }}\right)|\frac{-c_1 z_{\max }^2}{c_1 z_{\max }^2+1}\right)  \nonumber\\
&&+\left(\sqrt{4 c_1+1}+1\right) \Pi \left(\frac{2 c_1 z_{\max }^2}{\sqrt{4 c_1+1}-1};-\sin ^{-1}\left(\frac{z}{z_{\max }}\right)|\frac{-c_1 z_{\max }^2}{c_1 z_{\max }^2+1}\right) \Big],
\end{eqnarray} 
and $\theta_{>}(z)=\Omega-\theta_{<}(z)$ for $ \Omega/2\le \theta\le \Omega$, where $\Pi $ denotes the elliptic integral of the third kind. For region I with $\rho_*<0$, the opening angle is given by
\begin{eqnarray}\label{dual: 4d case opening angle negative T}
\Omega_{\text{I}}&=&2\theta_{<}(z_{\max })= \frac{-\sinh \left(\rho _*\right)}{\sqrt{\frac{c_1 \left(8 c_1+2\right)}{\sqrt{4 c_1 \text{sech}^2\left(\rho _*\right)+1}-1}}}\times \nonumber\\
&&\Big[ \left(\sqrt{4 c_1+1}-1\right) \Pi \left(\frac{1-\sqrt{4 c_1 \text{sech}^2\left(\rho _*\right)+1}}{\sqrt{4 c_1+1}+1}|\frac{2}{\sqrt{4 c_1 \text{sech}^2\left(\rho _*\right)+1}+1}-1\right)\nonumber\\
&&+\left(\sqrt{4 c_1+1}+1\right) \Pi \left(\frac{\sqrt{4 c_1 \text{sech}^2\left(\rho _*\right)+1}-1}{\sqrt{4 c_1+1}-1}|\frac{2}{\sqrt{4 c_1 \text{sech}^2\left(\rho _*\right)+1}+1}-1\right) \Big]
\end{eqnarray} 
where we have used (\ref{dual: 4d parameters}) above. Recall we have  $\Omega_{\text{II}}=\beta-\Omega_{\text{I}}(\rho_*\to -\rho_*)$ for region II with $\rho_*>0$. In the tensionless limit, we derive 
\begin{eqnarray}\label{dual: 4d opening angle small T limit}
\lim_{\rho_*\to 0}\Omega_{\text{I}}=\lim_{\rho_*\to 0}\Omega_{\text{II}}=\frac{ \sqrt{2} \pi }{\sqrt{\frac{\left(4 c_1+1\right) \left(\sqrt{4 c_1+1}-1\right)}{c_1}}}=\frac{\beta}{2}.
\end{eqnarray} 
It means $\Omega$ is a continuous function of $\rho_*$, which can be regarded as a check of our results. From (\ref{dual: 4d opening angle small T limit}), we solve
\begin{eqnarray}\label{dual: 4d f small T limit}
\lim_{\rho_*\to 0}f(\Omega)=c_1&=&\frac{-8 \Omega ^4+4 \pi ^2 \Omega ^2+\pi ^3 \sqrt{8 \Omega ^2+\pi ^2}+\pi ^4}{32 \Omega ^4},\\
&\approx& \begin{cases}
\frac{\pi ^4}{16 \Omega ^4},\  \ \ \ \ \ \ \ \ \ \text{for } \Omega \to 0,\\
\frac{2}{3 \pi } (\pi-\Omega),\  \ \text{for } \Omega \to \pi,
\end{cases} \label{dual: 4d f small two limits}
\end{eqnarray} 
which takes the expected limits (\ref{limits}) and gives $\lim_{\rho_*\to 0}\kappa_2=2/(3 \pi)$.  From \cite{Miao:2018dvm,Miao:2017aba}, we have $\lim_{\rho_*\to 0} C_D=120/\pi ^2$, which agrees with the universal relation (\ref{universal relation}). This is another support of our results. In section 3.3, we verify the universal relation (\ref{universal relation}) for general $d$ and $\rho_*$. 

{\bf Case III: $\rho_*=0$}

As discussed above, the calculations can be highly simplified in the tensionless case $\rho_*=0$. In such limit, we have by symmetry
\begin{eqnarray}\label{zero tension: opening angle}
\Omega_{\text{I}}=\Omega_{\text{II}}=\frac{\beta}{2}=\frac{2\pi z_{\max}}{d+(2-d) z_{\max}^2},
\end{eqnarray} 
where we have used $\lim_{\rho_* \to 0}z_h=z_{\max}$ from $c_1=(1-z_h^2)/z_h^d=(\text{sech}^2(\rho_*)-z_{\max}^2)/z_{\max}^d$. From (\ref{dual: EOM Q1 c1}), we have
\begin{eqnarray}\label{zero tension: f Omega0}
\lim_{\rho_*\to 0}f(\Omega)=c_1=z_{\max }^{-d} \left(1-z_{\max }^2\right).
\end{eqnarray} 
The above two equations yield
\begin{eqnarray}\label{zero tension: f Omega}
\lim_{\rho_*\to 0}f(\Omega)&=&\Omega ^d \left(\frac{\sqrt{(d-2) d \Omega ^2+\pi ^2}-\pi }{d-2}\right)^{-d} \left(1-\frac{\left(\pi -\sqrt{(d-2) d \Omega ^2+\pi ^2}\right)^2}{(d-2)^2 \Omega ^2}\right)\\
&\approx& \begin{cases}
d^{-d}(2 \pi )^d\frac{1}{ \Omega ^d},\ \ \ \ \ \ \ \ \text{for } \Omega \to 0,\\
\frac{2 }{\pi (d-1)}(\pi-\Omega ),\  \ \ \ \ \text{for } \Omega \to \pi,
\end{cases} \label{zero tension:  d two limits}
\end{eqnarray}
which yields $\lim_{\rho_*\to 0}\kappa_2=2/((d-1) \pi)$.  From \cite{Miao:2018dvm}, we read off $\lim_{\rho_*\to 0} C_D=2^{d+2} \pi ^{-\frac{d}{2}-\frac{1}{2}} \Gamma \left(\frac{d+3}{2}\right)$, which obeys the universal relation (\ref{universal relation}). We verify the universal relation (\ref{universal relation}) for general $\rho_*$ in the next subsection. 

\subsection{Two typical limits}

Now we consider two typical limits (\ref{limits}) of opening angles, where the integral (\ref{dual: opening angle result 1}) can be performed exactly. We verify that $\kappa_1$ is given by the coefficient (\ref{Tij strip}) of holographic strip in the small angle limit, and  $\kappa_2$ obeys the universal relation (\ref{universal relation}) in the smooth limit. These are strong supports of our results. 

{\bf Limit I: $\Omega\to 0$}

Performing a transformation $z=y \ z_{\max}$, we rewrite the integral (\ref{dual: opening angle result 1}) as
\begin{eqnarray}\label{limit I: opening angle integral}
\Omega_{\text{I}}=\int_0^1 dy \frac{2 z_{\max }}{\left(1-a y^d-\left(y^2-y^d\right) z_{\max }^2\right) \sqrt{\frac{a \left(y^d-1\right)+\left(y^2-y^d\right) z_{\max }^2}{a-1}}},
\end{eqnarray} 
where $a=\text{sech}^2\left(\rho _*\right)$. In the small angle limit, we have $f(\Omega)\to \kappa_1/\Omega^d\to \infty$, which yields $z_{\max}\to 0$ from (\ref{dual: EOM Q1 c1}).  In such limit, we obtain
\begin{eqnarray}\label{limit I: c1 limit}
f(\Omega)=c_1=\frac{a}{z_{\max}^d}+O(\frac{1}{z_{\max}^{d-1}}),
\end{eqnarray} 
and 
\begin{eqnarray}\label{limit I: opening angle integral 1}
\Omega_{\text{I}}&=&\frac{2 z_{\max}}{\sqrt{\frac{a}{1-a}}} \int_0^1 dy  \frac{1}{\sqrt{1-y^d} \left(1-a y^d\right)}+O(z_{\max}^2),\nonumber\\
&=& \frac{2 z_{\max}}{\sqrt{\frac{a}{1-a}}}\frac{\Gamma \left(\frac{1}{2}\right) \Gamma \left(\frac{1}{d}\right)}{d \Gamma \left(\frac{1}{2}+\frac{1}{d}\right)} \, _2F_1\left(\frac{1}{d},1;\frac{d+2}{2 d};a\right)+O(z_{\max}^2),
\end{eqnarray} 
and 
\begin{eqnarray}\label{limit I: opening angle integral 2}
\Omega_{\text{II}}
&=&\frac{4 \pi  z_{\max }}{a^{1/d} d }- \frac{2 z_{\max}}{\sqrt{\frac{a}{1-a}}}\frac{\Gamma \left(\frac{1}{2}\right) \Gamma \left(\frac{1}{d}\right)}{d \Gamma \left(\frac{1}{2}+\frac{1}{d}\right)} \, _2F_1\left(\frac{1}{d},1;\frac{d+2}{2 d};a\right)+O(z_{\max}^2).
\end{eqnarray} 
In the above derivations, we have used the integral formula
\begin{eqnarray}\label{limit I: integral formula}
\int_0^1dy \frac{1}{\sqrt{1-y^d} \left(1-a y^d\right)}=\frac{\Gamma \left(\frac{1}{2}\right) \Gamma \left(\frac{1}{d}\right)}{d \Gamma \left(\frac{1}{2}+\frac{1}{d}\right)} \, _2F_1\left(\frac{1}{d},1;\frac{d+2}{2 d};a\right),
\end{eqnarray} 
which can be derived from the standard integral expression of hypergeometric function by taking $t=y^d$
\begin{eqnarray}\label{limit I: integral formula Hyper}
\, _2F_1\left(a,b;c;z\right)=\frac{\Gamma(c)}{\Gamma(b)\Gamma(c-b)}\int_0^1dt \ t^{b-1}(1-t)^{c-b-1}(1-t z)^{-a}.
\end{eqnarray} 
From (\ref{limit I: c1 limit}, \ref{limit I: opening angle integral 1},\ref{limit I: opening angle integral 2}) and $a=\text{sech}^2\left(\rho _*\right)$, we obtain
\begin{eqnarray}\label{limit I: kappa1}
\kappa_{1}=\begin{cases}
\left(\frac{2 \Gamma \left(\frac{1}{2}\right) \Gamma \left(\frac{1}{d}\right) \left(-\tanh \left(\rho _*\right)\right) \text{sech}^{\frac{2}{d}-1}\left(\rho _*\right) \, _2F_1\left(1,\frac{1}{d};\frac{1}{d}+\frac{1}{2};\text{sech}^2\left(\rho _*\right)\right)}{d \Gamma \left(\frac{1}{2}+\frac{1}{d}\right)}\right){}^d, \ \ \ \ \ \text{for } \rho_*\le 0,\\
\left(\frac{4 \pi }{d}-\frac{2 \Gamma \left(\frac{1}{2}\right) \Gamma \left(\frac{1}{d}\right) \tanh \left(\rho _*\right) \text{sech}^{\frac{2}{d}-1}\left(\rho _*\right) \, _2F_1\left(1,\frac{1}{d};\frac{1}{d}+\frac{1}{2};\text{sech}^2\left(\rho _*\right)\right)}{d \Gamma \left(\frac{1}{2}+\frac{1}{d}\right)}\right){}^d,\  \ \text{for }  \rho_*\ge 0.
\end{cases} 
\end{eqnarray} 
As discussed in the introduction, $\kappa_1$ is related to the Casimir effects of strips. The holographic strip is studied in  \cite{Fujita:2011fp}, which yields the same $\kappa_1$ as above. It is a strong support to our proposal of holographic wedges.

{\bf Limit II: $\Omega\to \pi$}

Let us go on to discuss the smooth limit $\Omega\to \pi$. We have 
\begin{eqnarray}\label{limit II: c1 limit}
f(\Omega)=c_1=2 a^{\frac{1}{2}-\frac{d}{2}} \left(\sqrt{a}-z_{\max }\right)+O\left(\sqrt{a}-z_{\max }\right)^2,
\end{eqnarray} 
and
\begin{eqnarray}\label{limit II: opening angle integral 1}
\Omega_{\text{I}}&=&\int_0^1 dy \frac{2 \sqrt{\frac{1-a}{1-y^2}}}{1-a y^2}+O\left(\sqrt{a}-z_{\max }\right)^2
\nonumber\\
&+&\int_0^1dy\frac{2 a \left(\frac{1-a}{a \left(1-y^2\right)}\right)^{3/2} \left(y^d-1-a \left(3 y^{d+2}-2 y^d-2 y^4+y^2\right)\right) }{(1-a) \left(1-a y^2\right)^2} \left(\sqrt{a}-z_{\max }\right) \nonumber\\
&=&\pi +O\left(\sqrt{a}-z_{\max }\right)^2 \nonumber\\
&+&\frac{2 \sqrt{\pi } \Gamma \left(\frac{d+1}{2}\right) \left(a (1-a)^{-\frac{d}{2}} \, _2F_1\left(\frac{d-1}{2},\frac{d}{2};\frac{d+2}{2};\frac{a}{a-1}\right)-\frac{d}{\sqrt{1-a}}\right)}{\sqrt{a} d \Gamma \left(\frac{d}{2}\right)}\left(\sqrt{a}-z_{\max }\right)
\end{eqnarray}
and
\begin{eqnarray}\label{limit II: opening angle integral 2}
\Omega_{\text{II}}&=&\beta-\Omega_{\text{I}}=\pi+O\left(\sqrt{a}-z_{\max }\right)^2
\nonumber\\
&+&\Big( 2 \pi  (1-d) a^{\frac{1}{2}-\frac{d}{2}}-\frac{2 \sqrt{\pi } \Gamma \left(\frac{d+1}{2}\right) \left(a (1-a)^{-\frac{d}{2}} \, _2F_1\left(\frac{d-1}{2},\frac{d}{2};\frac{d+2}{2};\frac{a}{a-1}\right)-\frac{d}{\sqrt{1-a}}\right)}{\sqrt{a} d \Gamma \left(\frac{d}{2}\right)}\Big)\left(\sqrt{a}-z_{\max }\right).\nonumber\\
\end{eqnarray}
In the above calculations, we have used the integral formula \footnote{One can prove this formula by applying Mathematica together with various identities of hypergeometric functions for general $d$. Note that Mathematica cannot calculate directly the definite integral (\ref{limit II: integral formula}) for general $d$. The trick is that one first performs the indefinite integration by using Mathematica. }
\begin{eqnarray}\label{limit II: integral formula}
&&\int_0^1 dy \frac{2 a \left(\frac{1-a}{a \left(1-y^2\right)}\right)^{3/2} \left(y^d-1-a \left(3 y^{d+2}-2 y^d-2 y^4+y^2\right)\right)}{(1-a) \left(a y^2-1\right)^2}\nonumber\\
&=& \frac{2 \sqrt{\pi } \Gamma \left(\frac{d+1}{2}\right) \left(a (1-a)^{-\frac{d}{2}} \, _2F_1\left(\frac{d-1}{2},\frac{d}{2};\frac{d+2}{2};\frac{a}{a-1}\right)-\frac{d}{\sqrt{1-a}}\right)}{\sqrt{a} d \Gamma \left(\frac{d}{2}\right)}.
\end{eqnarray}
From (\ref{limit II: c1 limit},\ref{limit II: opening angle integral 1},\ref{limit II: opening angle integral 2}) together with $a=\text{sech}^2\left(\rho _*\right)$, we derive
\begin{eqnarray}\label{limit II: kappa2}
\kappa_{2}=\begin{cases}
\frac{-d \Gamma \left(\frac{d}{2}\right) \text{sech}^3\left(\rho _*\right) \left(-\text{csch}\left(\rho _*\right)\right){}^{-d}}{\sqrt{\pi } \Gamma \left(\frac{d+1}{2}\right) \left(\text{sech}^3\left(\rho _*\right) F+d \text{csch}\left(\rho _*\right) \left(-\tanh \left(\rho _*\right)\right){}^d\right)}, \ \ \ \ \ \ \ \ \ \ \ \ \ \ \ \ \ \ \ \ \text{for } \rho_*\le 0,\\
\frac{d \Gamma \left(\frac{d}{2}\right)}{\pi  (d-1) d \Gamma \left(\frac{d}{2}\right)-\frac{\sqrt{\pi } \Gamma \left(\frac{d+1}{2}\right) \left(d \left(\tanh \left(\rho _*\right) \text{sech}\left(\rho _*\right)\right){}^d-\sinh \left(\rho _*\right) \text{sech}^{d+3}\left(\rho _*\right) F\right)}{\text{sech}^2\left(\rho _*\right) \tanh ^{d+1}\left(\rho _*\right)}},\ \text{for }  \rho_*\ge 0,
\end{cases}
\end{eqnarray} 
where $F=\, _2F_1\left(\frac{d-1}{2},\frac{d}{2};\frac{d+2}{2};-\text{csch}^2\left(\rho _*\right)\right)$.
We remark that $\kappa_2$ is continuous at $\rho_*=0$. In fact, $\kappa_2$ with $\rho_*\ge 0$ and $\rho_*\le 0$ can be derived from the analytical continuations of each others.  Take $d=4$ as an example, (\ref{limit II: kappa2}) yields
\begin{eqnarray}\label{limit II: kappa2 d=4}
\kappa_{2}=\frac{2}{3 \pi \left(1+ \tanh \left(\rho _*\right)\right) },\ \ \ \ \text{for } d=4,
\end{eqnarray} 
for both $\rho_*\le 0$ and $\rho_*\ge 0$. $C_D$ for AdS/BCFT can be found in \cite{Miao:2018dvm}. We verify that $\kappa_{2}$ (\ref{limit II: kappa2},\ref{limit II: kappa2 d=4}) and $C_D$ of \cite{Miao:2018dvm} indeed obeys the universal relation (\ref{universal relation}), which is a non-trivial test of our results.

\begin{figure}[t]
\centering
\includegraphics[width=7.5cm]{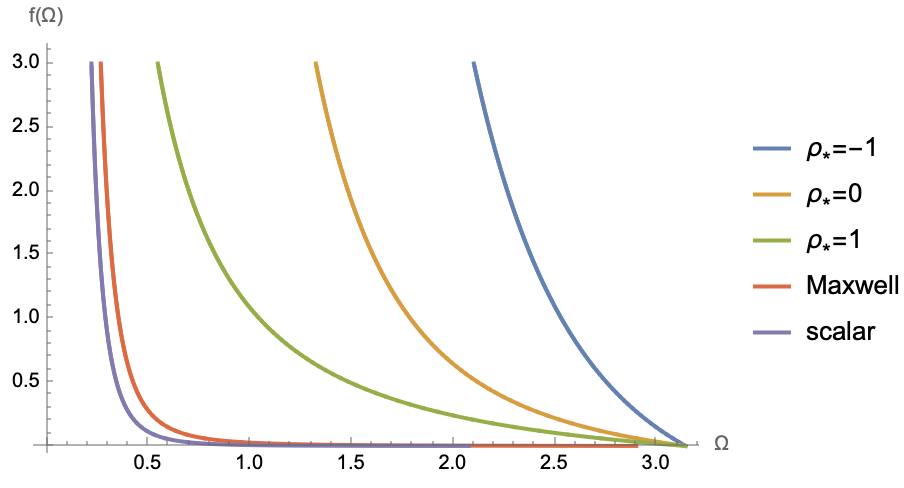} \includegraphics[width=7.5cm]{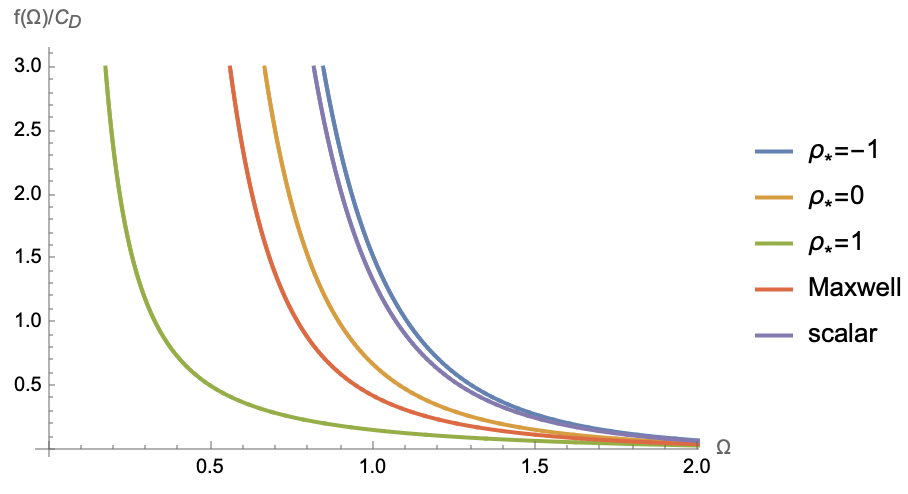}
\caption{Wedge Casimir effects of various 4d BCFTs. The left figure is for $f(\Omega)$ and the right figure is for  $f(\Omega)/C_D$. The blue, orange, green, purple and red curves denotes the Casimir effects for holographic BCFTs with $(\rho_*=-1, 0, 1)$, Maxwell theory and free scalar.  The left figure suggests that $f_{\rho_*=-1}>f_{\rho_*=0}>f_{\rho_*=1}>f_{\text{Maxwell}}>f_{\text{scalar}}$. The right figure implies that $f_{\rho_*=-1}/C_D>f_{\text{scalar}}/C_D>f_{\rho_*=0}/C_D>f_{\text{Maxwell}}/C_D>f_{\rho_*=1}/C_D$. }
\label{4dwedgeCasimir}
\end{figure}

To summarize, we construct the holographic dual of wedges and show that it yields the expected Casimir effects in this section. In particular, it obeys the universal relation (\ref{universal relation}), which strongly supports our results. To end this section, we draw the wedge Casimir effects for various 4d BCFTs in Fig. \ref{4dwedgeCasimir}.  It shows that $f'(\Omega)\le 0$, which agrees with the discussions of section 2.3. Besides, it implies that the holographic wedge Casimir energy increases with the brane tension $T=(d-1) \tanh(\rho_*)$. In particular, the negative tension (smaller tension) tends to produce larger $f(\Omega)$ and $f(\Omega)/C_D$, or equivalently, smaller Casimir energy density $T_{tt}=-f(\Omega)/r^d$. It is natural that the more negative the brane tension, the more negative the Casimir energy density. 

\section{Holographic wedge entanglement}

This section studies the holographic entanglement entropy (HEE) of the subsystem $0\le r\le L$ in a wedge \footnote{This paper studies the entanglement entropy for BCFT living in a wedge. Please distinguish the entanglement entropy between a wedge region and its complement for CFTs. The latter case has been studied in \cite{Bueno:2019mex} with interesting results.}. The RT surface generally depends on the angle $\theta$, complicating the calculations. For the tensionless case $\rho_*=0$, we have $\Omega_{\text{I}}=\Omega_{\text{II}}=\beta/2$ (\ref{zero tension: opening angle}) by symmetry. It means that regions I and II, as well as their bulk duals and RT surfaces, are symmetric. Thus, the RT surface of a wedge with opening angle $\beta/2$ is just half that of the whole space and is $\theta$-independent. We focus on this case and leave the tensive case for future works. For $\rho_*=0$, all the non-trivial contributions to entanglement entropy come from the corner of the wedge. We call such contribution wedge entanglement, which vanishes when the wedge disappears $\Omega=\pi$. We find that wedge entanglement increases with the opening angle $\Omega$, similar to the wedge Casimir energy. 

We focus on the bulk metric (\ref{dual: metric}) with $0\le \theta \le \Omega=\beta/2$ and the following ansatz of RT surface
\begin{eqnarray}\label{EE: RT}
t=\text{constant}, \ r=F(z).
\end{eqnarray} 
Then, the area of the RT surface reads
\begin{eqnarray}\label{EE: RT area}
A=\Omega \int_{\epsilon}^{z_h} \frac{1}{z^{d-1} F(z)^{d-3}} \sqrt{1+\frac{h(z) F'(z)^2}{F(z)^2}} \ dz ,
\end{eqnarray} 
where $\epsilon$ is the UV cutoff and $z_h$ is defined by $h(z_h)=0$.  We have set the tangential volume $V=\int d^{d-3}y=1$ for simplicity. Taking variations of (\ref{EE: RT area}) with respect to $F(z)$, we derive the equation of motion (EOM) 
\begin{eqnarray}\label{EE: EOM}
F''(z)&=&-\frac{(d-3) F(z)}{h(z)} -\frac{(d-4) F'(z)^2}{F(z)} \nonumber\\
&+&\frac{F'(z)^3 \left(2 (d-1) h(z)-z h'(z)\right)}{2 z F(z)^2}+\frac{F'(z) \left((d-1) h(z)-z h'(z)\right)}{z h(z)}.
\end{eqnarray} 

\subsection{AdS$_4$/BCFT$_3$}

\begin{figure}[t]
\centering
\includegraphics[width=10cm]{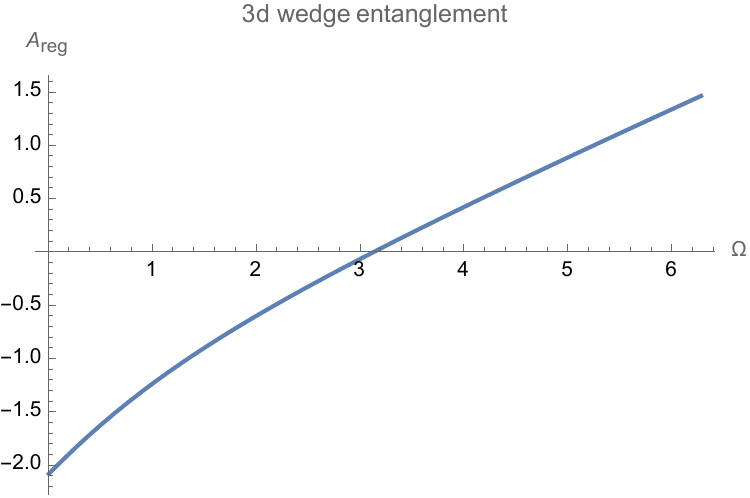} 
\caption{ Wedge contribution to entanglement entropy for $d=3$. It shows the renormalized wedge EE $A_{\text{reg}}/(4 G_N)$ increases with the opening angle $\Omega$. }
\label{3dwedgeEE}
\end{figure}

Let us first study the case $d=3$, for which we can make analytical discussions. When $d=3$, there is an exact solution to EOM (\ref{EE: EOM})
\begin{eqnarray}\label{EE: 3d solution}
r=F(z)=L,
\end{eqnarray} 
where $L$ denotes the scale of the subregion $0\le r \le L$. Substituting (\ref{EE: 3d solution}) into the area functional (\ref{EE: RT area}), we get 
\begin{eqnarray}\label{EE: RT area 1}
A&=&\Omega \Big(\frac{1}{\epsilon}-\frac{1}{z_h} \Big)\nonumber\\
&=&\Omega  \left(\frac{1}{\epsilon }+\frac{\Omega }{\pi -\sqrt{3 \Omega ^2+\pi ^2}}\right),
\end{eqnarray} 
where we have used (\ref{dual: period}) and $\Omega=\beta/2$ in the above calculations.  The divergent term labels the area law of entanglement entropy. We are interested in the wedge contributions to entanglement entropy (EE), so we deduct the plane EE with $ z_h=1$,
\begin{eqnarray}\label{EE: RT area 0}
A_0=\int_0^{\Omega} d\theta \Big(\frac{1}{\epsilon}-1 \Big).
\end{eqnarray}
 Recall that for Casimir energy of parallel plates, we subtract the vacuum contribution between the parallel plates $\int_0^L dx...$ rather than the whole space $\int_{-\infty}^{\infty} dx...$. Thus, above,  we deduct  $\int_0^{\Omega} d\theta...$ instead of $\int_0^{\pi} d\theta...$.  In this way, we get a finite renormalized area
\begin{eqnarray}\label{EE: RT area reg}
A_{\text{reg}}=A-A_0=\Omega \Big(1-\frac{1}{z_h} \Big)=\Omega  \left(1+\frac{\Omega }{\pi -\sqrt{3 \Omega ^2+\pi ^2}}\right).
\end{eqnarray} 
The renormalized EE is given by $A_{\text{reg}}/(4 G_N)$. See Fig. \ref{3dwedgeEE}, which shows that the renormalized wedge EE increases with the opening angle $\Omega$. Note that the renormalized wedge EE can be negative generally. Like the Casimir energy, nothing goes wrong for a renormalized physical quantity to be negative.  We stress that the non-renormalized wedge EE (\ref{EE: RT area 1}) is positive and increases with $\Omega$ too. 

\subsection{AdS$_{d+1}$/BCFT$_d$}

Let us go on to study the case $d>3$. By the shooting method, we can solve EOM (\ref{EE: EOM}) numerically. We show only vital points below. See \cite{Hu:2022zgy} for more technical details, where a similar problem (the island in AdS/dCFT) has been studied. Here, we have different motivations.  

Solving EOM (\ref{EE: EOM}) perturbingly around $z=z_h$, we get
\begin{eqnarray}\label{EE: perturbative solution}
F(z)=L_0+ b_1 (z-z_h)+ b_2 (z-z_h)^2+...
\end{eqnarray} 
where 
\begin{eqnarray}\label{EE: perturbative solution bi}
b_1=-\frac{(d-3)  z_h}{(d-2) z_h^2-d} L_0, \ b_2=-\frac{(d-3) ((d-14) d+29)  z_h^2}{8 \left(d-(d-2) z_h^2\right){}^2}L_0.
\end{eqnarray} 
We define a new function $F_0(z)=L_0+ b_1 (z-z_h)+ b_2 (z-z_h)^2$ with $b_i$ given by (\ref{EE: perturbative solution bi}). Then, the initial values become
\begin{eqnarray}\label{EE: initial values}
F(z_h-\epsilon_1)=F_0(z_h-\epsilon_1), \  F'(z_h-\epsilon_1)=F'_0(z_h-\epsilon_1),
\end{eqnarray} 
where $\epsilon_1$ is a small cutoff for numeral calculations. For any given $L_0, z_h$, we can numerically solve EOM (\ref{EE: EOM}). We adjust the input $L_0$ to satisfy the boundary condition on the AdS boundary $z=\e$
 \begin{eqnarray}\label{EE: L L0}
F(\e)= L.
\end{eqnarray} 
It is the so-called shooting method. Without loss of generality, we set $L=1$ below.

\begin{figure}[t]
\centering
\includegraphics[width=7.5cm]{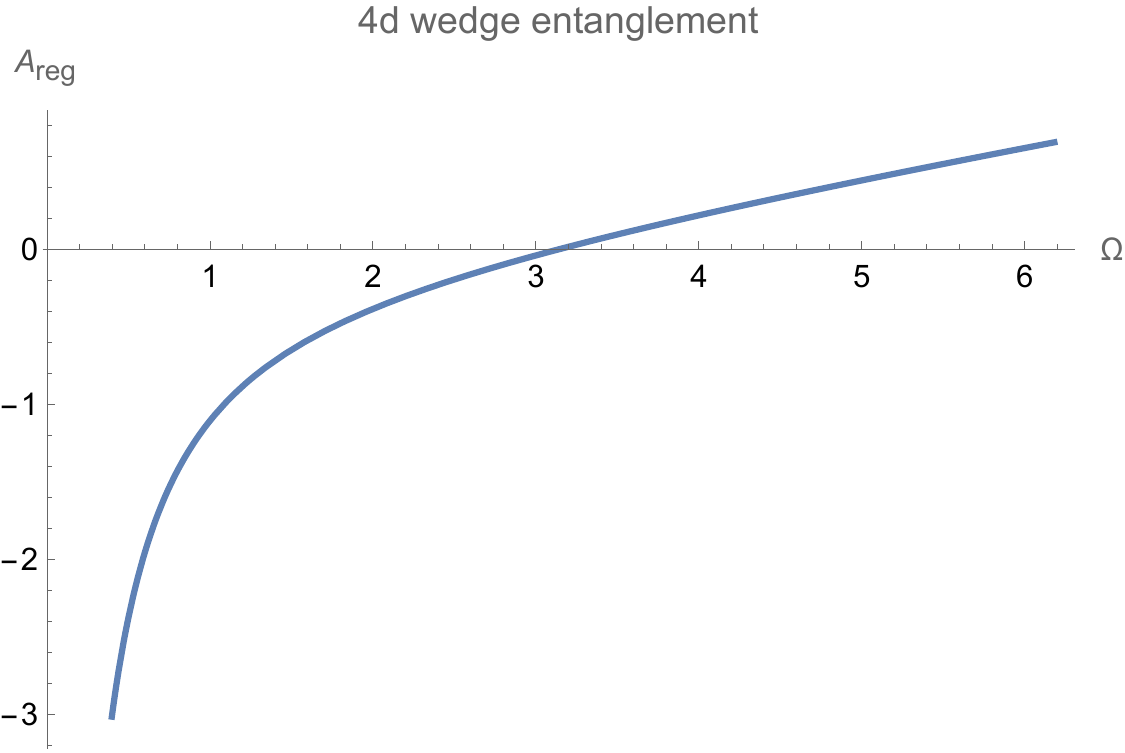} \includegraphics[width=7.5cm]{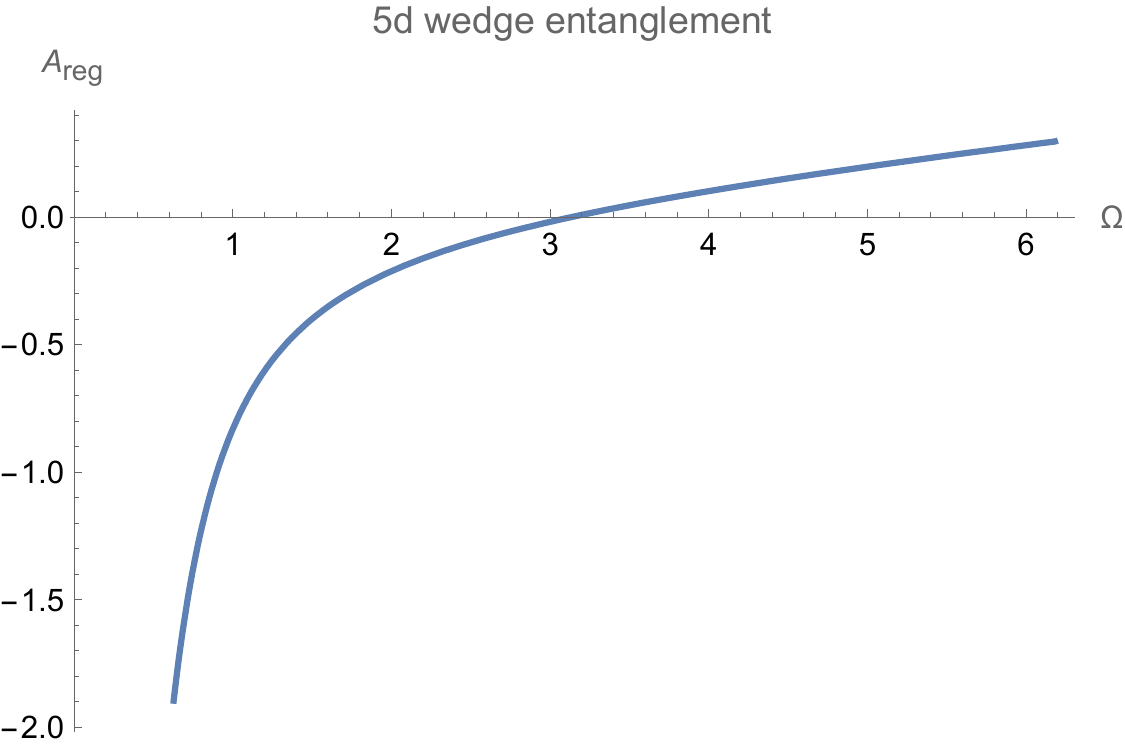} 
\caption{ Wedge contribution to entanglement entropy for $d=4$ (left) and $d=5$ (right). It shows the renormalized wedge EE $A_{\text{reg}}/(4 G_N)$ increases with the opening angle $\Omega$. }
\label{4d5dwedgeEE}
\end{figure}

Similar to the above subsection, we can obtain the wedge EE by deducting the plane EE. See Fig. \ref{4d5dwedgeEE}, which shows the renormalized wedge EE increases with the opening angle $\Omega$. We remark that the Casimir energy density of wedges $-f(\Omega)/r^d$ also increases with $\Omega$ since $f'(\Omega)<0$. It is interesting to explore if some relations existed between the wedge Casimir effect and wedge EE. We leave it to future works. 

\section{Toward holographic polygons}

This section generalizes the discussions to polygons. For simplicity, we focus on AdS$_3$/BCFT$_2$ so that we can employ the power of conformal mappings. In two dimensions ($d=2$), we can conformally map the upper half-plane to the interior of a plane polygon. Unfortunately, one cannot do so for polytopes or ``polygons"  \footnote{Similar to wedges, by ``polygons" in higher dimensions, we mean the direct product of a 2-dimensional polygon and $(d-2)$-dimensional flat space, i.e., $P^2\times R^{d-2}$. } in higher-dimensional flat space. Generally, one gets a ``polygon" in curved space instead of flat space for $d>2$. We leave the non-trivial problems in higher dimensions to future works and focus on $d=2$ in this paper. We hope it could shed some light on the holographic constructions of polytopes and ``polygons" in higher dimensions.   

By making the Schwarz–Christoffel mapping, 
\begin{eqnarray}\label{polygon: SC mapping}
w=h^{-1}(\xi)=\int^{\xi} \frac{c_0}{(s-a_1)^{1-\alpha_1/\pi}(s-a_2)^{1-\alpha_2/\pi}...(s-a_{n-1})^{1-\alpha_{n-1}/\pi}} ds,
\end{eqnarray} 
one can map the upper half-plane $\text{Im}(\xi)\ge 0$ to the interior of a plane n-polygon, where $\alpha_i$ are interior angles of the polygon, and $a_i$ are real numbers related to the vertices of the polygon.  The above conformal mappings correspond to the following coordinate transformations in bulk 
\begin{eqnarray}\label{polygon: coordinate transformations1}
&&\xi=h(w)-\frac{2z^2 (h')^2\bar{h}''}{4|h'|^2+z^2 |h''|^2}, \\ \label{polygon: coordinate transformations2}
&&\bar{\xi}=\bar{h}(\bar{w})-\frac{2z^2 (\bar{h}')^2h''}{4|h'|^2+z^2 |h''|^2},\\
&&\eta=\frac{4z (h'\bar{h}')^{3/2}}{4|h'|^2+z^2 |h''|^2}. \label{polygon: coordinate transformations3}
\end{eqnarray} 
Under these coordinate transformations, the Poincar\'e AdS$_3$ 
\begin{eqnarray}\label{polygon: P AdS3}
ds^2=\frac{d\eta^2+d\xi d\bar{\x}}{\eta^2},
\end{eqnarray} 
becomes 
\begin{eqnarray}\label{polygon: new AdS3}
ds^2=\frac{dz^2+\Big(1+z^4 T(w)\bar{T}(\bar{w})\Big)dw d\bar{w}+ z^2 T(w) dw^2+ z^2 \bar{T}(\bar{w}) d\bar{w}^2}{z^2},
\end{eqnarray} 
where $T(w) $ and $\bar{T}(\bar{w})$ are chiral and anti-chiral energy stress tensors
\begin{eqnarray}\label{polygon: T barT}
T(w)=\frac{3 (h'')^2-2 h' h'''}{4 h'^2}, \ \  \bar{T}(\bar{w})=\frac{3 (\bar{h}'')^2-2 \bar{h}' \bar{h}'''}{4 \bar{h}'^2}.
\end{eqnarray} 
Note that, according to (\ref{dual: Tij holo formu}), the holographic stress tensor reads $T_{ww}=2 T(w)$.  Similarly, the EOW brane becomes 
\begin{eqnarray}\label{polygon: P Q}
\frac{\xi-\bar{\xi}}{2 i}=-\sinh(\rho_*) \eta,
\end{eqnarray} 
where $(\xi, \bar{\xi}, \eta)$ are given by (\ref{polygon: coordinate transformations1},\ref{polygon: coordinate transformations2},\ref{polygon: coordinate transformations3}).  Let us study some examples below.  

{\bf Case I: semi-infinite strip}

\begin{figure}[t]
\centering
\includegraphics[width=10cm]{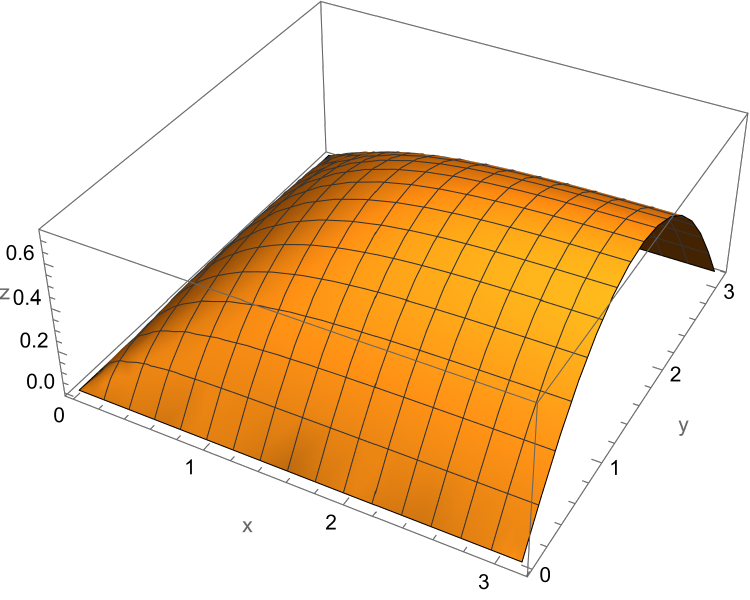}
\caption{AdS$_3$/BCFT$_2$ for semi-infinite strip with $\rho_*=-1$. The EOW brane is a smooth surface in bulk and the singularity appears only at the corner of semi-infinite strip on the AdS boundary.}
\label{2dhalfstrip}
\end{figure}

The semi-infinite strip can be considered a special triangle limit with the interior angles $(\pi/2, \pi/2,0)$. The corresponding Schwarz–Christoffel mapping is
\begin{eqnarray}\label{polygon: semi-infinite strip}
w=h^{-1}(\xi)=\int^{\xi} \frac{1}{(s-1)^{1/2}(s+1)^{1/2}} ds=\cosh ^{-1}(\xi),
\end{eqnarray} 
where we have chosen suitable integral constant so that $h^{-1}(1)=0, h^{-1}(-1)=i \pi, h^{-1}(\infty)=\infty$. From (\ref{polygon: semi-infinite strip}), we get $h(w)=\cosh(w)$, which yields 
\begin{eqnarray}\label{polygon: T barT strip }
T(w)&=&\frac{1}{4} \left(1+3 \text{csch}^2(w)\right)\nonumber\\
&\approx&\begin{cases}
\frac{3}{4 w^2}, \ \ \ \text{for }w \to 0,\\
\frac{1}{4}, \ \ \ \ \ \  \text{for }w \to \infty.
\end{cases}
\end{eqnarray} 
The above stress tensor has the expected limits near the corner and at infinity.  
Transforming into the coordinates $(r, \theta)$ with $w=r \exp(i \theta)$, we derive near the corner $w\to 0$
\begin{eqnarray}\label{polygon: Trr strip }
T_{rr}\approx \frac{3}{r^2},
\end{eqnarray} 
which agrees with the stress tensor (\ref{dual: f 2d}) of 2d wedge with $\Omega=\pi/2$. Similarly, 
 in coordinates $(x, y)$ with $w=x+ i y$, we have at infinity $w\to \infty$
 \begin{eqnarray}\label{polygon: Txx strip}
T_{xx}\approx 1,
\end{eqnarray} 
which agrees with the strip stress tensor $\kappa_1/L^2$ with $L=\pi$ and $\kappa_1=\pi$ (\ref{limit I: kappa1}).  We draw the EOW brane (\ref{polygon: P Q}) in Fig. \ref{2dhalfstrip}, which is a smooth surface in bulk. 

{\bf Case II: triangle}

\begin{figure}[t]
\centering
\includegraphics[width=7.5cm]{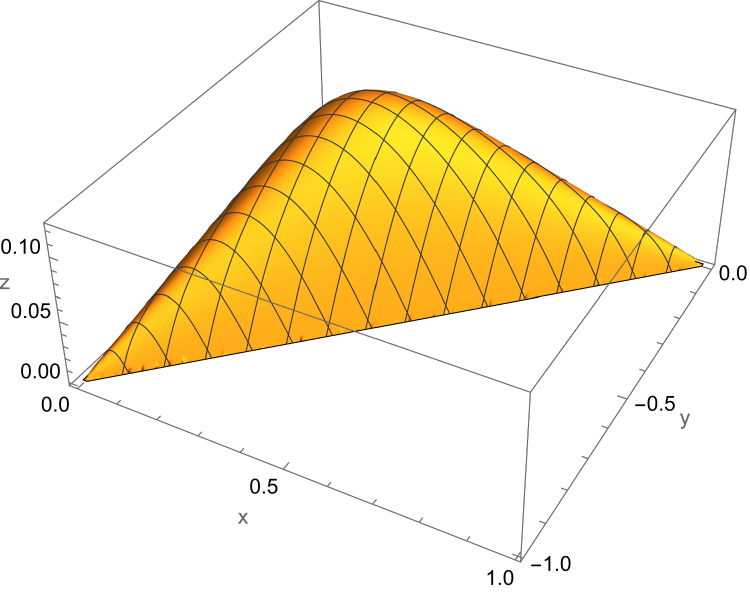} \includegraphics[width=7.5cm]{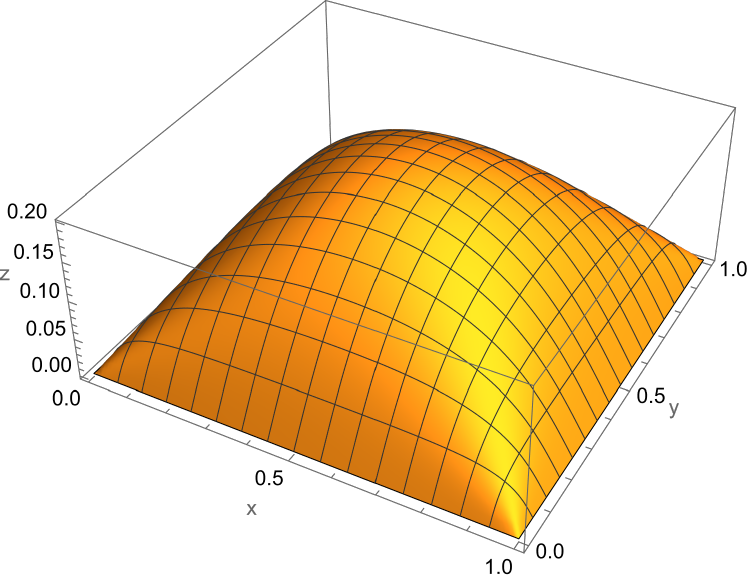} 
\caption{AdS$_3$/BCFT$_2$ for triangle (left) and square (right) with $\rho_*=-1$. The interior angles of triangle are $(\frac{\pi}{2}, \frac{\pi}{4},  \frac{\pi}{4})$. These figures show that the EOW brane is a smooth surface in bulk and the singularity appears only at the corner of the polygon on the AdS boundary.}
\label{2dtrianglesquare}
\end{figure}

For the triangle with interior angles $\alpha, \beta$ and  $\pi-\alpha-\beta$, the conformal mapping is 
\begin{eqnarray}\label{polygon: triangle}
w&=&h^{-1}(\xi)=\int^{\xi} \frac{c_0}{(s-1)^{1-\alpha/\pi}(s+1)^{1-\beta/\pi}} ds\nonumber\\
&=&c_0\frac{\pi  2^{\frac{\beta }{\pi }-1} (\xi-1)^{\alpha /\pi } \, _2F_1\left(\frac{\alpha }{\pi },1-\frac{\beta }{\pi };\frac{\alpha +\pi }{\pi };\frac{1-\xi}{2}\right)}{\alpha },
\end{eqnarray} 
where we have chosen a suitable integral constant and the parameter 
\begin{eqnarray}\label{polygon: triangle c0}
c_0=\frac{\Gamma \left(1-\frac{\beta }{\pi }\right) \Gamma \left(\frac{\alpha +\beta }{\pi }\right)}{\pi  e^{i \alpha } 2^{\frac{\alpha +\beta -\pi }{\pi }} \Gamma \left(\frac{\alpha }{\pi }\right) \csc (\beta )},
\end{eqnarray} 
so that $h^{-1}(1)=0, h^{-1}(-1)=1, h^{-1}(\infty)=e^{-i \alpha } \sin (\beta ) \csc (\alpha +\beta )$. Unfortunately, one cannot get the exact expression of $h(w)$ but only its inverse function $h^{-1}(\xi)$. By using $h'(w)=1/(h^{-1})'(\xi)$ and so on, we rewrite (\ref{polygon: T barT}) as
\begin{eqnarray}\label{triangle: T barT}
T(w)=\frac{-3 ((h^{-1})'')^2+2 (h^{-1})' (h^{-1})'''}{4 (h^{-1})'^4}, \ \  \bar{T}(\bar{w})=\frac{-3 ((\bar{h}^{-1})'')^2+2 (\bar{h}^{-1})' (\bar{h}^{-1})'''}{4 (\bar{h}^{-1})'^4}.
\end{eqnarray} 
Near the corner $\xi\to1, w\to 0$, we derive from (\ref{polygon: triangle},\ref{triangle: T barT})
\begin{eqnarray}\label{triangle: T near corner0}
w=(\xi -1)^{\alpha /\pi } \left(\frac{2^{-\frac{\alpha }{\pi }} e^{-i \alpha } \Gamma \left(\frac{\alpha +\beta }{\pi }\right) \Gamma \left(1-\frac{\beta }{\pi }\right) \sin (\beta )}{\alpha  \Gamma \left(\frac{\alpha }{\pi }\right)}+O\left((\xi -1)^1\right)\right),
\end{eqnarray}
and
\begin{eqnarray}\label{triangle: T near corner}
T(w)&=&(\xi -1)^{-\frac{2 \alpha }{\pi }} \left(\frac{2^{\frac{2 \alpha }{\pi }-2} e^{2 i \alpha } \left(\pi ^2-\alpha ^2\right) \csc ^2(\beta ) \Gamma \left(\frac{\alpha }{\pi }\right)^2}{\Gamma \left(\frac{\alpha +\beta }{\pi }\right)^2 \Gamma \left(1-\frac{\beta }{\pi }\right)^2}+O\left((\xi -1)^1\right)\right)\nonumber\\
&=& \frac{\pi ^2-\alpha ^2}{4 \alpha ^2 w^2}+O(\frac{1}{w}),
\end{eqnarray} 
where we have used (\ref{triangle: T near corner0}) to derive the second line of (\ref{triangle: T near corner}).  Changing into the coordinates $(r, \theta)$, we get near the corner
\begin{eqnarray}\label{triangle: T near corner 1}
T_{rr}= \frac{\pi ^2-\alpha ^2}{\alpha ^2 r^2}+O(\frac{1}{r}),
\end{eqnarray} 
which exactly agrees with the holographic stress tensor (\ref{dual: f 2d}) of 2d wedge. It is a double-check of our results. From (\ref{polygon: P Q}) and (\ref{polygon: triangle}), we can numerically obtain the embedding function of EOW brane. See Fig. \ref{2dtrianglesquare} (left). The cases of quadrangles and polygons are similar to that of triangles. One can obtain their holographic duals following the above approach. For simplicity, we do not repeat the calculations. See Fig. \ref{2dtrianglesquare} (right) for an example of the holographic square. 

In summary, in this section, we construct the holographic duals of 2d polygons by applying the Schwarz-Christoffel mappings and their bulk correspondences. We verify that the holographic stress tensors approach those of wedges near the corners of polygons. Besides, the EOW branes dual to polygons are smooth surfaces in bulk. It is similar to the RT surface of a corner, which is also a smooth surface in bulk \cite{Hirata:2006jx,Myers:2012vs}. From NBC (\ref{NBC}), we see that the extrinsic curvatures of the EOW brane are proportional to the induced metrics. Since the induced metrics are continuous functions, so are the extrinsic curvatures. It means that the EOW brane is smoothly embedded into the bulk. Thus, it must be a smooth surface.  

To end this section, let us comment on the constructions of holographic ``polygons" in higher dimensions. First, the holographic stress tensor should approach that of a wedge near the corner of the ``polygon". Second, the EOW brane is a smooth surface in bulk, and the singularity appears only at the corner of the ``polygon" on the AdS boundary. Third, due to the non-trivial Casimir effect, the gravity dual of the ``polygon" cannot be Poincar\'e AdS, even not locally AdS. Recall the example of holographic wedges in section 3, where the bulk geometry is not an AdS.

\section{Conclusions and Discussions}

This paper investigates the Casimir effect of the wedge and its holographic dual. We are interested in the universal properties of the wedge Casimir effect for BCFTs. We prove a universal relation between the displacement operator and wedge Casimir effect in the smooth limit, i.e., $\Omega\to \pi$. It should be stressed that this universal relation applies to general states of BCFTs in general curved wedge space for $r\to 0$ and $\Omega\to \pi$. For instance, it works for $r\ll 1/T, r\ll 1/\sqrt{R}$ for a thermal state in curved space, where $r$ is the proper distance to the corner of the wedge, $T$ is the temperature and $R$ denotes the curvature scale. Besides, we find that the wedge Casimir energy increases with the opening angle, i.e., $f'(\Omega)\le 0$, and verify it with several examples. Next, we construct the holographic dual of wedges in the initial model of AdS/BCFT. We observe that it cannot be the Poincar\'e AdS due to the non-trivial Casimir effect of wedges. 
Instead, it is given by a cutout of the AdS soliton.
We verify that our proposal can produce the correct smooth and singular limits: it obeys the universal relation between the displacement operator and Casimir effect for $\Omega\to \pi$; it reduces to the holographic strip for $r\to \infty, \Omega\to 0$ with fixed $r \Omega=L$. These are strong supports of our results. Next, we discuss the wedge contribution to holographic entanglement entropy.
Interestingly, the renormalized wedge entanglement entropy increases with the opening angle, similar to the wedge Casimir energy. Finally, we study the holographic polygon in AdS$_3$/BCFT$_2$ and briefly discuss its generalization to higher dimensions. The important lesson we learned is that the EOW brane is a smooth surface in bulk due to the continuous extrinsic curvatures determined by the Neumann boundary condition.

Let us discuss several interesting problems worthy of further exploration. 

First, we notice that the wedge Casimir effect is similar to the so-called corner entanglement \cite{Bueno:2015rda, Bueno:2015xda, Miao:2015dua} in several aspects. Since the wedge becomes a corner for $d=3$, we take 3d CFT as an example. The entanglement entropy of 3d CFT takes the following form in the ground state
\begin{eqnarray}\label{conclusion: 3dEE}
S=B l/\epsilon- a(\Omega) \ln (l/\epsilon)+O(1),
\end{eqnarray} 
where $a(\Omega)$ characterizes the underlying CFT and plays a similar role of $f(\Omega)$ of this paper.  $a(\Omega)$ obeys $a(\Omega)=a(2\pi-\Omega)$ \footnote{Entanglement entropy obeys $S(A)=S(\bar{A})$ for a pure state, where $\bar{A}$ is the complement of subsystem $A$. Thus, we have $a(\Omega)=a(2\pi-\Omega)$ for a pure state, which is different from $f(\Omega)$ of wedge Casimir effect.}, and behaves as 
\begin{eqnarray}\label{conclusion:limits}
a(\Omega)\to \begin{cases}
k_1/\Omega,\  \ \ \ \ \ \ \ \ \text{for } \Omega \to 0,\\
k_2 (\pi-\Omega)^2,\  \ \text{for } \Omega \to \pi.
\end{cases}
\end{eqnarray}
In particular, $k_2$ obeys the universal relation 
\begin{eqnarray}\label{conclusion:universal relation}
k_2=\frac{\pi^2}{24} C_T,
\end{eqnarray}
where $C_T$ is the central charge defined by two-point functions of the stress tensor. Remarkably, $a(\Omega)$ obeys similar limits (\ref{conclusion:limits}) and universal relation (\ref{conclusion:universal relation}) as (\ref{limits},\ref{universal relation}) of wedge Casimir effect. It implies a mysterious relationship between entanglement entropy and the Casimir effect may exist, which is worth further study. One possible is that they are related by the Weyl anomaly. Thus, it is also interesting to derive the Weyl anomaly in general wedge space.

Second, the corner entanglement obeys the inequalities \cite{Hirata:2006jx, Casini:2008as}
\begin{eqnarray}\label{conclusion: da dda}
a(\Omega)\ge 0, \  a'(\Omega)\le 0, \ a''(\Omega)\le 0, \ \ \ \text{for} \ 0 \le \Omega\le \pi,
\end{eqnarray}
which can be derived from the strong subadditivity of entanglement entropy. We argue that $f'(\Omega)\le 0$ by applying the virtual displacement method in this paper. We have checked that $f''(\Omega)\ge 0$ is obeyed by all the examples studied in this paper. Getting a strict proof of $f''(\Omega)\ge 0$ is an interesting problem. Inspired by the corner entanglement, if strong-subadditivity-like inequality exists for the Casimir effect, it is also an interesting problem.  

Third, there are two natural generalizations of the results of this paper. One is the holographic dual of ``polygon" and polyhedron in higher dimensions, as mentioned in section 5. The other one is the holographic dual of a cone 
\begin{eqnarray}\label{conclusion: cone}
ds^2=dr^2 + r^2 d\Omega_m^2-dt^2+\sum_{a=1}^{d-m-2} dy_a^2,
\end{eqnarray}
where $ d\Omega_m^2$ denotes the line element of unit sphere $S^m$ with $m\ge 2$.  

Fourth, this paper focuses on pure Einstein gravity for simplicity. It is interesting to generalize to the case of matter fields and higher derivative gravity. Besides, we focus on the holographic dual of the ground state of the wedge. Studying the gravity duals of the thermal and excited states is interesting.

Last, this paper focuses on the case with the same boundary conditions on the two boundaries $\theta=0,\Omega$ of the wedge. What happens if we impose different boundary conditions? In addition to the geometric singularity, there is a new singularity from the discontinuous boundary conditions at the corner of the wedge. It is interesting to explore the holographic dual of this case.

\section*{Acknowledgements}

We thank Chong-Sun Chu for valuable discussions on the general problem of constructing AdS/BCFT with fixed BCFT geometry. We are grateful to Jian-Xin Lu for the helpful comments and discussions. This work is supported by the National Natural Science Foundation of China (No. 12275366 and No. 11905297).

\appendix

\section{Wedge Casimir effect for free scalars}

\subsection{Heat kernel method}

This appendix investigates the Casimir effect of wedges in general dimensions by applying the heat kernel method. Unfortunately, we only obtain the results for even dimensions. We leave the discussions for odd $d$ to the next appendix. 

We are interested in the conformally coupled scalar with the action
\begin{eqnarray}\label{scalaraction}
I=\frac{1}{2}\int_{M}d^dx\sqrt{|g|} (\nabla_i \phi \nabla^i \phi+\xi R \phi^2)+\int_{\partial M} d^{d-1}y\sqrt{|h|} \xi k \phi^2,
\end{eqnarray} 
where $\xi=\frac{d-2}{4(d-1)}$ and $k$ is the trace of extrinsic curvatures.  We impose the following conformally invariant boundary conditions 
\begin{equation}\label{BCscalar}
\begin{split}
&\text{Dirichlet BC} : \phi|_{\partial M}=0,\\
&\text{Robin BC} : \ \ (\nabla_n + 2\xi k)\phi|_{\partial M}=0.
\end{split}
\end{equation}
The heat kernel satisfies the equation of motion (EOM)
 \begin{eqnarray}\label{EOM heat kernel}
\partial_s K(s,x_i,x'_i)-(\nabla^j\nabla_j-\xi R) K(s,x_i,x'_i)=0,
\end{eqnarray}
 and the limit
 \begin{eqnarray}\label{BC heat kernel}
\lim_{s\to 0} K(s,x_i,x'_i)=\frac{\delta^{(d)}(x_i-x'_i)}{\sqrt{|g|}}. 
\end{eqnarray}
It also obeys the BC (\ref{BCscalar}) on the boundary of the wedge. From the heat kernel, we can obtain the Green function
 \begin{eqnarray}\label{Greenfunction heat kernel}
G(x_i,x'_i)=\int_0^{\infty} ds K(s,x_i,x'_i),
\end{eqnarray}
and then derive the renormalized stress tensor by
\begin{eqnarray}\label{Tij heat kernel}
\langle T_{ij} \rangle=\lim_{x'_i\to x_i} \left[  (1-2\xi) \nabla_{i}\nabla_{j'}-2\xi \nabla_{i} \nabla_{j} +(2\xi -\frac{1}{2})g_{ij} \nabla_{l} \nabla^{l'}  +\xi (R_{ij}+\frac{4\xi-1}{2}Rg_{ij}) \right] \hat{G}(x_i,x'_i),\nonumber\\
\end{eqnarray}
where $\hat{G}=G-G_0$ with $G_0$ the Green function in the free space without boundaries.

For the flat wedge space (\ref{metric}) $W=R^{d-2}\times C^2$, the heat kernel is given by 
\begin{eqnarray}\label{heat kernel}
K(s,x_i,x'_i)=K_{R^{d-2}} K_{C^2}
\end{eqnarray}
where \cite{Vassilevich:2003xt}
\begin{eqnarray}\label{heat kernel R}
K_{R^{d-2}}=\frac{1}{(4\pi s)^{\frac{d-2}{2}}} \exp(-\frac{\sum_{\hat{i}=1}^{d-2} (x_{\hat{i}}-x'_{\hat{i}})^2}{4s})
\end{eqnarray}
and for DBC and (Robin boundary condition) RBC \cite{Deutsch:1978sc}
\begin{eqnarray}\label{heat kernel C}
K_{C^2}=\int_{0}^{\infty} dp \frac{2 p}{\Omega }  \exp \left(-s p^2 \right)J_{\frac{m \pi }{\Omega }}(p r) J_{\frac{m \pi }{\Omega }}\left(p r'\right)  \begin{cases} \sum_{m=1}^{\infty} \sin \left(\frac{ m \pi \theta }{\Omega }\right) \sin \left(\frac{m \pi \theta '}{\Omega }\right),\\
\sum_{m=0}^{\infty} \cos \left(\frac{m \pi \theta }{\Omega }\right) \cos \left(\frac{m \pi \theta '}{\Omega }\right).
\end{cases}
\end{eqnarray}
Substituting (\ref{heat kernel}) into (\ref{Greenfunction heat kernel}) and applying the integral formula  \cite{Deutsch:1978sc}
\begin{eqnarray}\label{heat kernel formula}
\int_{0}^{\infty} dp p^{-\lambda} J^2_{v}\left(p r\right) =\frac{r^{\lambda-1} \Gamma(\lambda) \Gamma(v+\frac{1}{2}-\frac{\lambda}{2})}{2^{\lambda} \Gamma^2(\frac{1}{2}+\frac{\lambda}{2}) \Gamma(v+\frac{1}{2}+\frac{\lambda}{2})},
\end{eqnarray}
we obtain 
\begin{eqnarray}\label{heat kernel G limit}
\lim_{x'\to x} G=\frac{\pi ^{1-\frac{d}{2}}  \Gamma (3-d) \Gamma \left(\frac{d}{2}+\frac{m \pi }{\Omega }-1\right)}{r^{d-2}\Omega  \Gamma \left(2-\frac{d}{2}\right) \Gamma \left(-\frac{d}{2}+\frac{m \pi }{\Omega }+2\right)} \begin{cases} \sum_{m=1}^{\infty} \sin^2 \left(\frac{ m \pi \theta }{\Omega }\right)\\
\sum_{m=0}^{\infty} \cos^2 \left(\frac{m \pi \theta }{\Omega }\right)
\end{cases},
\end{eqnarray}
\begin{eqnarray}\label{heat kernel ddyG limit}
\lim_{x'\to x} \partial^2_{\hat{i}}G=-\frac{2 \pi ^{1-\frac{d}{2}} \Gamma (1-d) \Gamma \left(\frac{d}{2}+\frac{m \pi }{\Omega }\right)}{ r^d \Omega  \Gamma \left(1-\frac{d}{2}\right) \Gamma \left(-\frac{d}{2}+\frac{m \pi }{\Omega }+1\right)} \begin{cases} \sum_{m=1}^{\infty} \sin^2 \left(\frac{ m \pi \theta }{\Omega }\right)\\
\sum_{m=0}^{\infty} \cos^2 \left(\frac{m \pi \theta }{\Omega }\right)
\end{cases},
\end{eqnarray}
\begin{eqnarray}\label{heat kernel ddthetaG limit}
\lim_{x'\to x} \partial^2_{\theta}G=-\frac{\pi ^{3-\frac{d}{2}}  \Gamma (3-d) \Gamma \left(\frac{d}{2}+\frac{m \pi }{\Omega }-1\right)}{r^{d-2}\Omega ^3 \Gamma \left(2-\frac{d}{2}\right) \Gamma \left(-\frac{d}{2}+\frac{m \pi }{\Omega }+2\right)} \begin{cases} \sum_{m=1}^{\infty} m^2  \sin^2 \left(\frac{ m \pi \theta }{\Omega }\right)\\
\sum_{m=0}^{\infty} m^2  \cos^2 \left(\frac{m \pi \theta }{\Omega }\right)
\end{cases},
\end{eqnarray}
\begin{eqnarray}\label{heat kernel ddthetaG1 limit}
\lim_{x'\to x} \partial_{\theta} \partial_{\theta'}G=\frac{\pi ^{3-\frac{d}{2}}  \Gamma (3-d) \Gamma \left(\frac{d}{2}+\frac{m \pi }{\Omega }-1\right)}{r^{d-2}\Omega ^3 \Gamma \left(2-\frac{d}{2}\right) \Gamma \left(-\frac{d}{2}+\frac{m \pi }{\Omega }+2\right)} \begin{cases} \sum_{m=0}^{\infty} m^2 \cos^2 \left(\frac{m \pi \theta }{\Omega }\right)\\
\sum_{m=1}^{\infty} m^2  \sin^2 \left(\frac{ m \pi \theta }{\Omega }\right)
\end{cases}.
\end{eqnarray}
From the above four equations, we obtain the relations
\begin{eqnarray}\label{heat kernel relation1}
\lim_{x'\to x} \partial^2_{\hat{i}}G=\lim_{x'\to x} \frac{\frac{(d-2)^2}{4}G+\partial^2_{\theta}G}{r^2(1-d)},
\end{eqnarray}
and
\begin{eqnarray}\label{heat kernel relation2}
\lim_{x'\to x} \partial_{\theta} \partial_{\theta'} \begin{cases} G_{D}\\
G_{R}
\end{cases}=-\lim_{x'\to x}  \partial^2_{\theta} \begin{cases} G_{R}\\
G_{D}
\end{cases},
\end{eqnarray}
where $D$ and $R$ denotes DBC and RBC, respectively. So we can transform the derivatives of renormalized stress tensor (\ref{Tij heat kernel}) into the derivatives of $\theta$ due to the symmetries of wedges \cite{Deutsch:1978sc}.  From (\ref{Tij wedge}, \ref{Tij heat kernel}) together with (\ref{heat kernel ddthetaG limit},\ref{heat kernel relation1},\ref{heat kernel relation2}), we get the characteristic function
\begin{eqnarray}\label{characteristic function f}
f(\Omega)&=&\lim_{x'\to x} \frac{ r^{d-2}}{2(d-1)} \partial^2_{\theta}(G_{D}+G_{R}-2G_0)\nonumber\\
&=& \sum_{m=1}^{\infty}\frac{2^{-d} \pi ^{\frac{5}{2}-\frac{d}{2}} m^2 \Gamma \left(\frac{1}{2}-\frac{d}{2}\right) \Gamma \left(\frac{d}{2}+\frac{m \pi }{\Omega }-1\right)}{\Omega ^3 \Gamma \left(-\frac{d}{2}+\frac{m \pi }{\Omega }+2\right)} - (\Omega\to \pi). 
\end{eqnarray}
The above sum can be calculated directly for even $d$. For instance, we have for $d=2$
\begin{eqnarray}\label{heat kernel f2d}
f_{2d}(\Omega)=\sum_{m=1}^{\infty} -\frac{\pi  m}{2 \Omega ^2} - (\Omega\to \pi)= -\frac{\pi  \zeta (-1)}{2 \Omega ^2} - (\Omega\to \pi)=\frac{\frac{\pi ^2}{\Omega ^2}-1}{24 \pi },
\end{eqnarray}
which takes the expected form.  For $d=4$, we get
\begin{eqnarray}\label{heat kernel f4d}
f_{4d}(\Omega)=\sum_{m=1}^{\infty}\frac{\pi ^2 m^3}{12 \Omega ^4} - (\Omega\to \pi)= \frac{\pi ^2 \zeta(-3)}{12 \Omega ^4}  - (\Omega\to \pi)=\frac{\frac{\pi ^4}{\Omega ^4}-1}{1440 \pi ^2},
\end{eqnarray}
which agrees with the result of  \cite{Deutsch:1978sc}.  Similarly, we obtain
\begin{eqnarray}\label{heat kernel f6d}
f_{6d}(\Omega)=\frac{\pi \zeta(-3)}{120 \Omega ^4}-\frac{\pi ^3 \zeta(-5)}{120 \Omega ^6}  - (\Omega\to \pi)=\frac{-31 \Omega ^6+21 \pi ^4 \Omega ^2+10 \pi ^6}{302400 \pi ^3 \Omega ^6},
\end{eqnarray}
and 
\begin{eqnarray}\label{heat kernel f6d}
f_{8d}(\Omega)= \frac{\pi ^4 \zeta(-7)}{1680 \alpha ^8}-\frac{\pi ^2 \zeta(-5)}{336 \alpha ^6}+\frac{\zeta(-3)}{420 \alpha ^4} - (\Omega\to \pi)=\frac{-289 \Omega ^8+168 \pi ^4 \Omega ^4+100 \pi ^6 \Omega ^2+21 \pi ^8}{8467200 \pi ^4 \Omega ^8}.
\end{eqnarray}
Unfortunately, we do not derive results for odd $d$. We leave this problem to the next appendix.

\subsection{Mirror method}

In this appendix, we derive the wedge Casimir effect for the open angle $\Omega=\pi/m$ with an integer $m$, and then analytically extend the results to arbitrary open angle $\Omega$. 
By apply the mirror method, we obtain the Green functions for DBC and RBC
\begin{eqnarray} 
&&G_D=\frac{\Gamma(\frac{d}{2}-1)}{4 \pi^{\frac{d}{2}}} \sum_{i=1}^{2m} \frac{(-1)^{i+1}}{( r^2+r'^2-2 r r' \cos(\theta-\theta_i))^{\frac{d-2}{2}}}, \label{Green function D}\\
&&G_R=\frac{\Gamma(\frac{d}{2}-1)}{4 \pi^{\frac{d}{2}}} \sum_{i=1}^{2m} \frac{1}{( r^2+r'^2-2 r r' \cos(\theta-\theta_i))^{\frac{d-2}{2}}}, \label{Green function R}
\end{eqnarray}
where $\theta_i=(-1)^{i+1}\theta'+\frac{2\pi}{m} [\frac{i}{2}]$, and $[\ ]$ denotes the integer part. The background Green function in free space is given by
\begin{eqnarray} \label{Green function 0}
G_0=\frac{\Gamma(\frac{d}{2}-1)}{4 \pi^{\frac{d}{2}}}  \frac{1}{( r^2+r'^2-2 r r' \cos(\theta-\theta'))^{\frac{d-2}{2}}}.
\end{eqnarray}

From (\ref{characteristic function f}), we have
\begin{eqnarray} \label{Mirror: f}
f(\Omega)&=&\lim_{r'\to r, \theta'\to \theta} \frac{r^{d-2}}{2(d-1)} \partial_{\theta}^2 (G_D+G_R-2 G_0)\nonumber\\
&=&\frac{2^{-d-1} \pi ^{-\frac{d}{2}} \Gamma \left(\frac{d}{2}\right)}{d-1} \sum _{j=1}^{m-1} \csc^{d} \left(\frac{\pi  j}{m}\right)\left(d-1-(d-2) \sin^2\left(\frac{\pi  j}{m}\right)\right).
\end{eqnarray}
For even $d$, the above finite sum can be calculated by Mathematica. For instance, we have 
\begin{eqnarray} \label{Mirror: f even d}
f(\frac{\pi}{m})=\begin{cases}
\frac{m^2-1}{24 \pi }, \ \ \ \ \ \ \ \ \ \ \ \ \ \ \ \ \ \ \ \ \ \ \ \text{for d=2},\\
\frac{m^4-1}{1440 \pi ^2},\  \ \ \ \ \ \ \ \ \ \ \ \ \ \ \ \ \ \ \ \ \ \text{for d=4},\\
\frac{10 m^6+21 m^4-31}{302400 \pi ^3},\  \ \ \ \  \ \ \ \ \ \ \ \text{for d=6},\\
\frac{21 m^8+100 m^6+168 m^4-289}{8467200 \pi ^4}, \ \text{for d=8}.
\end{cases}
\end{eqnarray}
Performing the analytical continuation $\frac{\pi}{m}\to \Omega$, we get
\begin{eqnarray} \label{Mirror: f even d 1}
f(\Omega)=\begin{cases}
\frac{\frac{\pi ^2}{\Omega ^2}-1}{24 \pi }, \ \ \ \ \ \ \ \ \ \ \ \ \ \ \ \ \ \ \ \ \ \ \ \ \text{for d=2},\\
\frac{\frac{\pi ^4}{\Omega ^4}-1}{1440 \pi ^2},\  \ \ \ \ \ \ \ \ \ \ \ \ \ \ \ \ \ \ \ \ \ \ \text{for d=4},\\
\frac{\frac{10 \pi ^6}{\Omega ^6}+\frac{21 \pi ^4}{\Omega ^4}-31}{302400 \pi ^3},\  \ \ \ \  \ \ \ \ \ \ \ \ \ \text{for d=6},\\
\frac{\frac{21 \pi ^8}{\Omega ^8}+\frac{100 \pi ^6}{\Omega ^6}+\frac{168 \pi ^4}{\Omega ^4}-289}{8467200 \pi ^4}, \ \ \ \ \text{for d=8},
\end{cases}
\end{eqnarray}
which agree with the results of appendix A.1

The cases of odd $d$ are more subtle.  Take $d=3$ as an example, we need to calculate the sum
\begin{eqnarray} \label{Mirror: f 3d}
f_{3d}(\frac{\pi}{m})=\sum _{j=1}^{m-1} \left(\frac{\csc ^3\left(\frac{\pi  j}{m}\right)}{32 \pi }-\frac{\csc \left(\frac{\pi  j}{m}\right)}{64 \pi }\right).
\end{eqnarray}
Unfortunately, the sum formula for $\sum _{j=1}^{m-1} \csc^{2n+1}\left(\frac{\pi  j}{m}\right)$ ($n,m$ integers) is unknown.  The main idea is to replace the sum with an integral so that it is easier to perform the analytical continuation.  We have 
\begin{eqnarray} \label{Mirror: csc 1}
\csc(\pi s)= \int_0^{\infty } \frac{x^{s-1}}{\pi  (x+1)} \, dx, 
\end{eqnarray}
which yields
\begin{eqnarray} \label{Mirror: csc 2}
\sum_{j=1}^{m-1}\csc(\frac{\pi  j}{m})= \int_0^{\infty } \frac{x-x^{1/m}}{\pi  x (x+1) \left(x^{1/m}-1\right)} \, dx.
\end{eqnarray}
Taking two derivatives of (\ref{Mirror: csc 1}) with respect to $s$, we get
\begin{eqnarray} \label{Mirror: csc 3}
\partial_s^2\csc(\pi s)= 2 \pi ^2 \csc ^3(\pi  s)-\pi ^2 \csc (\pi  s)=\int_0^{\infty } \frac{x^{s-1} \log ^2(x)}{\pi  (x+1)} \, dx.
\end{eqnarray}
From (\ref{Mirror: csc 1},\ref{Mirror: csc 3}), we solve
\begin{eqnarray} \label{Mirror: csc 4}
\csc^3(\pi s)= \int_0^{\infty } \frac{x^{s-1} \left(\log ^2(x)+\pi ^2\right)}{2 \pi ^3 (x+1)} \, dx,  
\end{eqnarray}
which gives 
\begin{eqnarray} \label{Mirror: csc 5}
\sum_{j=1}^{m-1}\csc^3(\frac{\pi  j}{m})= \int_0^{\infty } \frac{\left(x-x^{1/m}\right) \left(\log ^2(x)+\pi ^2\right)}{2 \pi ^3 x (x+1) \left(x^{1/m}-1\right)} \, dx.
\end{eqnarray}
Substituting (\ref{Mirror: csc 2},\ref{Mirror: csc 5}) into (\ref{Mirror: f 3d}) and making analytical the continuation $m\to\pi/\Omega$, we obtain
\begin{eqnarray} \label{Mirror: f 3d result}
f_{3d}(\Omega)&=&\int_0^{\infty } \frac{\left(x-x^{\Omega /\pi }\right) \log ^2(x)}{64 \pi ^4 x (x+1) \left(x^{\Omega /\pi }-1\right)} \, dx\nonumber\\
&=& \int_0^{1} \frac{\left(x-x^{\Omega /\pi }\right) \log ^2(x)}{32 \pi ^4 x (x+1) \left(x^{\Omega /\pi }-1\right)} \, dx.
\end{eqnarray}
where we have used $\int_0^1...dx=\int_1^{\infty}...dx$ for the first line of integral.  The above equation has the correct smooth limit $f_{3d}(\pi)=0$, which is test of the result. At the next order, we have 
\begin{eqnarray} \label{Mirror: f 3d smooth limit}
f_{3d}(\Omega)&=&\int_0^1 \frac{(\pi -\Omega ) \log ^3(x)}{32 \left(\pi ^5 \left(x^2-1\right)\right)} \, dx+O(\Omega -\pi )^2\nonumber\\
&=&\frac{\pi -\Omega }{512 \pi }+O(\Omega -\pi )^2,
\end{eqnarray}
which yields $\kappa_2=1/(512\pi)$. Recall that we have $C_D=1/(8\pi^2)$ from (\ref{CD higher D scalar}). One can check that $\kappa_2$ and $C_D$ obeys the universal relation (\ref{universal relation}), which is another support of our results.  Following the above approach, we can derive $f(\Omega)$ for higher odd $d$ one by one. We list the results below. 
\begin{eqnarray} \label{Mirror: f 5d result}
f_{5d}(\Omega)&=&\int_0^{1} \frac{\left(x-x^{\Omega /\pi }\right) \log ^2(x) \left(\log ^2(x)+\pi ^2\right)}{1024 \pi ^7 x (x+1) \left(x^{\Omega /\pi }-1\right)} \, dx\nonumber\\
&=& \frac{3 (\pi -\Omega )}{16384 \pi ^2}+O(\Omega -\pi )^2.
\end{eqnarray}
\begin{eqnarray} \label{Mirror: f 7d result}
f_{7d}(\Omega)&=&\int_0^{1}  \frac{\left(x-x^{\Omega /\pi }\right) \log ^2(x) \left(\log ^2(x)+\pi ^2\right) \left(\log ^2(x)+9 \pi ^2\right)}{49152 \pi ^{10} x (x+1) \left(x^{\Omega /\pi }-1\right)}  \, dx\nonumber\\
&=& \frac{25 (\pi -\Omega )}{524288 \pi ^3}  +O(\Omega -\pi )^2.
\end{eqnarray}
We verify that the above $\kappa_2$ obeys the universal relation (\ref{universal relation}) for $d=5,7$ too. See discussions above Table. \ref{table1kappa2}. 

\section{Poincar\'e AdS}

In this section, we show that Poincar\'e AdS space cannot be the dual of a wedge. For simplicity, we focus on Poincar\'e AdS$_3$
\begin{eqnarray} \label{appB: AdS}
ds^2=\frac{dz^2+dr^2+r^2 d\theta^2}{z^2}, \ \ 0\le \theta \le \Omega.
\end{eqnarray}
From scaling symmetry of AdS and the fact that there is no other scale in the problem, we take the following ansatz of the EOW brane 
\begin{eqnarray} \label{appB: EOW brane}
z=r F(\theta),
\end{eqnarray}
where $F(\theta)$ obeys the boundary conditions
\begin{eqnarray} \label{appB: brane BC}
F(0)=F'(\frac{\Omega}{2})=0.
\end{eqnarray}
We remark that (\ref{appB: EOW brane}) is inspired from the RT surface of wedge \cite{Hirata:2006jx,Myers:2012vs}. Similar to section 3, we focus on region I with negative tension below. By considering its complement, we automatically get region II with positive tension. See Fig. \ref{wedgeAdSBCFT}.  Substituting (\ref{appB: EOW brane}) into NBC (\ref{NBC}), we get one independent equation
\begin{eqnarray} \label{appB: NBC equation}
\frac{1}{\sqrt{F'(\theta )^2+F(\theta )^2+1}}+\tanh \left(\rho _*\right)=0,
\end{eqnarray} 
which can be solved as 
\begin{eqnarray} \label{appB: NBC solution}
F(\theta )=\pm\text{csch}\left(\rho _*\right) \sin (\theta +c_0).
\end{eqnarray} 
Imposing the boundary conditions (\ref{appB: brane BC}), we obtain a constraint for the opening angle of wedge
\begin{eqnarray} \label{appB: NBC solution 1}
\Omega=\pi,
\end{eqnarray} 
which means the wedge becomes a plane. In other words, Poincar\'e AdS$_3$ is not dual to a wedge but a half space. The above discussions can be generalized to higher dimensions straightforwardly. 
To end this section, we mention that the metric (\ref{dual: metric 2d}) of section 3 is dual to a wedge. The embedding function of EOW brane can be read off from (\ref{dual: coordinate transformation 2d z},\ref{dual: 2d case angle}).

\end{document}